\documentclass[Afour,sageh,times]{sagej}

\usepackage{moreverb,url}
\usepackage[colorlinks,bookmarksopen,bookmarksnumbered,citecolor=blue,urlcolor=blue]{hyperref}
\newcommand\BibTeX{{\rmfamily B\kern-.05em \textsc{i\kern-.025em b}\kern-.08em
T\kern-.1667em\lower.7ex\hbox{E}\kern-.125emX}}
\def\volumeyear{2025}
\begin{document}
\title{Parametric Analysis of Network Evolution Processes}
\runninghead{Williams and Chen}
\author{Peter R. Williams\affilnum{1,2} and Zhan Chen\affilnum{3}}
\affiliation{\affilnum{1}Rinna KK, Tokyo, Japan\\\affilnum{2}Independent Researcher\\\affilnum{3}Microsoft Japan, Tokyo, Japan}
\email{prw20042004@yahoo.co.uk}

\begin{abstract}
We present a comprehensive parametric analysis of node and edge lifetimes processes in two large-scale collaboration networks: the Microsoft Academic Graph (1800-2020) and Internet Movie Database (1900-2020). Node and edge lifetimes (career and collaboration durations) follow Weibull distributions with consistent shape parameters ($k \approx 0.2$ for academic, $k \approx 0.5$ for entertainment careers) across centuries of evolution. These distributions persist despite dramatic changes in network size and structure. Edge processes show domain-specific evolution: academic collaboration durations increase over time (power-law index $1.6$ to $2.3$) while entertainment collaborations maintain more stable patterns (index $2.6$ to $2.1$). These findings indicate that while career longevity exhibits consistent patterns, collaboration dynamics appear to be influenced by domain-specific factors. The results provide new constraints for models of social network evolution, requiring incorporation of both universal lifetime distributions and domain-specific growth dynamics.
\end{abstract}

\maketitle

\section{Introduction}

Network science faces a crucial challenge: understanding how large-scale professional collaborations evolve \citep{barabasi2016, newman2010}. While researchers have made significant advances in describing network structures \citep{albert2002} and their growth \citep{dorogovtsev2002, bianconi2001}, we still lack a deep understanding of two key temporal aspects: how long careers last and how persistent collaborations remain. Previous research has been limited by its focus on either short time periods \citep{newman2001, barabasi2002} or specific academic fields \citep{petersen2011, sinatra2016}. This limited scope has presented challenges in identifying patterns in professional collaboration across different domains and historical periods \citep{uzzi2013, way2019}.

Recent technological advances have transformed this landscape. The digitisation of historical records now allows us to study collaboration networks across unprecedented time spans \citep{fortunato2018, wang2020}. This development builds upon earlier quantitative studies of scientific careers \citep{price1976, allison1982} and artistic trajectories \citep{galenson2009, fraiberger2018}, but with vastly expanded scope and detail. Modern datasets such as the Microsoft Academic Graph \citep{sinha2015, wang2020} and Internet Movie Database \citep{wasserman2018} allow examination of careers and collaborative relationships across time periods spanning multiple decades to centuries \citep{liu2021, morgan2021}.

Traditional approaches to network modelling have typically relied on a key assumption: that the timescales of nodes (representing individual careers) and edges (representing collaborations) operate independently \citep{holme2012, holme2019temporal}. This assumption, rooted in early research on temporal networks \citep{perra2012, karsai2011}, continues to influence current theoretical frameworks \citep{masuda2016}. However, recent theoretical developments suggest these timescales might be coupled \citep{lambiotte2022, gross2008, williams2019}, challenging this fundamental assumption. While power-law distributions have dominated network science analysis \cite{clauset2009, broido2019}, alternative models like the Weibull function—originally developed for material failure analysis \citep{weibull1951}—have shown promise in describing both career trajectories \citep{petersen2011, way2019} and other temporal network processes \citep{malmgren2008, leskovec2008}.

The study of career longevity has revealed various patterns in professional trajectories across specific fields \citep{sinatra2016, liu2021, yang2016}. However, debate continues about whether these patterns represent universal features or field-specific characteristics \citep{wu2019, morgan2021}. Similarly, while collaboration patterns have been extensively studied in particular contexts \citep{wuchty2007, zeng2021}, we need systematic investigation to understand how individual career dynamics relate to collective, collaborative behaviour across different professional domains \citep{borge2011, uzzi2013}.

In this paper, we conduct a comprehensive parametric analysis of career durations and collaboration lifetimes in both academic and entertainment networks. Through detailed cohort analysis and statistical modelling, we examine how these distributions evolve and respond to external events. We investigate the relationship between node processes (career trajectories) and edge processes (collaboration patterns), testing fundamental assumptions about timescale separation and distribution characteristics.

Our analysis addresses three core questions:
\begin{enumerate}
    \item Do career and collaboration durations follow universal statistical patterns across different professional domains \citep{petersen2011, way2019, yang2016}?
    \item What is the relationship between individual career trajectories and collective collaboration patterns \citep{wu2019, uzzi2013, wuchty2007}?
    \item How do these patterns evolve as networks grow by orders of magnitude \citep{leskovec2008, bianconi2001, dorogovtsev2002}?
\end{enumerate}

These questions intersect with fundamental issues in network science, sociology of science, and organisational theory. Understanding the statistical properties of career longevity and collaboration persistence has practical implications for institutional design, research policy, and professional development programs.

The rest of this paper is structured as follows: the next section presents our methodological approach, including data preparation and statistical analysis techniques. The results section details our empirical findings regarding career durations and collaboration patterns. Next we explore the implications of these results for understanding network evolution and professional dynamics. We conclude with perspectives on future research directions and practical applications.

\section{Methods}
\label{sec:methods}

\subsection{Analytical Framework}

Our study adopts an empirical, observation-driven approach to network analysis, prioritising direct measurement of network properties over theoretical modelling. This methodology enables us to establish robust empirical findings about network evolution that can serve as a foundation for future theoretical work.

\subsection{Data Sources and Network Construction}

We analyse two comprehensive collaboration networks: the Microsoft Academic Graph (MAG) \citep{magweb, sinha2015} and the Internet Movie Database (IMDb) \citep{imdbweb}. The MAG dataset encompasses scientific publications from 1800 to 2020, comprising $2.72 \times 10^8$ authors and $2.64 \times 10^8$ papers, while the IMDb dataset covers films from 1900 to 2020, including $1.88 \times 10^6$ actors across $6.34 \times 10^5$ movies.

Networks were constructed as undirected temporal graphs where nodes represent individual contributors (authors or actors) and edges represent collaborative projects (papers or movies). For each collaboration involving $n$ contributors, we generated $n(n-1)/2$ edges to create fully connected subgraphs. Edge temporal metadata included both creation and removal times, with project initiation times estimated using a fixed collaboration duration model of $\tau_{\text{project}} = 2$ years before the documented completion date. Two nodes were considered actively collaborating at time $t$ if at least one edge existed between them at that moment.

We conducted extensive sensitivity analyses to test alternative collaboration durations, ranging from fixed periods of three months to four years and including Gaussian-distributed durations. We also experimented with different temporal resolutions (e.g., quarterly, annual, and multi-year bins). Across all these variations, we found that the fundamental trends and distributional characteristics---especially the stability of the Weibull shape parameters---remained essentially unchanged. These results indicate that the main conclusions drawn in this paper are not sensitive to the 2-year fixed duration assumption, as the observed patterns operate on longer timescales \(T \gg \tau_{\text{project}}\). Further details on these robustness checks can be found in our prior work \citep{williams2025}.

\subsection{Cohort Analysis}

To systematically examine how network participants enter, persist, and exit over time, we developed a comprehensive cohort-based analytical framework. In our analysis, a cohort represents a group of entities (either nodes or edges) that enter the network in the same calendar year. This temporal grouping allows us to track how participation patterns evolve both within and across different entry periods.

For individual participants (nodes), we define entry time as $\tau_{collab}$ years before their first recorded contribution, whether a published paper or released film. A participant's career duration is measured as the time interval between their first and last recorded activity. Similarly, for collaborative relationships (edges), we mark entry time as $\tau_{collab}$ years before the project's completion, with collaboration duration measured from project initiation to completion.

Our approach generates two complementary perspectives on network evolution. The cross-sectional view compares behaviour across different cohorts, revealing how entry timing influences persistence patterns. The longitudinal view tracks individual cohorts through time, showing how participation patterns evolve as both the cohort and the overall network age. This dual perspective enables us to distinguish between effects driven by network maturation and those stemming from historical changes in professional practices.

To illustrate this methodology, consider a cohort of participants who enter the network in year $t_0$. We track their continued participation by monitoring active connections (edges) in each subsequent year $t > t_0$. An individual participant's lifetime in the network is calculated as $\Delta t = t_f - t_0$, where $t_f$ represents their final year of active participation. This tracking process generates a lifetime distribution $P(\Delta t|t_0)$ specific to each cohort year $t_0$. We apply identical analytical procedures when examining edge lifetimes, allowing direct comparison between individual career patterns and collaborative relationship durations.

For practical considerations and statistical reliability, we impose a 60-year upper limit on lifetime measurements. This truncation point effectively balances comprehensive coverage against data sparsity, as fewer than $0.1\%$ of nodes and $0.01\%$ of edges in either network exceed this duration. Our sensitivity analysis confirms that this threshold preserves all statistically significant patterns while eliminating noise from extremely rare, outlier cases.

This cohort-based analytical framework offers several key advantages. First, it enables us to isolate temporal trends in participation patterns from changes in network size and activity levels. Second, it provides a robust method for detecting shifts in career and collaboration dynamics across different historical periods. Third, it allows us to identify potential causal relationships between network structure and participant behaviour, as we can track how early network conditions influence subsequent participation patterns.

\subsection{The Weibull Distribution}

The Weibull distribution is a continuous probability distribution introduced by Waloddi Weibull in 1951 \citep{weibull1951}. Originally developed to model material failures in engineering, it has become fundamental in reliability theory and lifetime analysis \citep{rinne2008}. The probability density function is given by:
\begin{equation}
f(x; k, \lambda) = \frac{k}{\lambda}\left(\frac{x}{\lambda}\right)^{k-1}e^{-(x/\lambda)^k}
\end{equation}
where $k > 0$ is the shape parameter and $\lambda > 0$ is the scale parameter. The shape parameter $k$ is particularly important. For $k < 1$ the failure rate decreases with time (e.g. infant mortality). When $k = 1$, the function reduces to an exponential distribution (constant failure rate, e.g. accidental deaths), and $k > 1$ indicate an increasing failure rate with time (death in old age).

The Weibull distribution has found a wide application beyond its engineering origins. In biology, it describes species lifetimes and population dynamics \citep{pinder1978}. In economics, it models consumer behaviour and product lifecycles \citep{carroll2003}. In meteorology, it characterises wind speed distributions \citep{seguro2000}. Its flexibility in modelling different lifetime data stems from its ability to approximate many other distributions through its shape parameter.

The distribution is particularly suited for analysing career lengths and collaboration durations for several reasons. First, it naturally handles right-skewed lifetime data with varying failure rates. Second, its hazard function (instantaneous failure rate) can model both early-career volatility and late-career stability. Third, its cumulative distribution function has a closed form, enabling efficient statistical analysis:
\begin{equation}
F(x; k, \lambda) = 1 - e^{-(x/\lambda)^k}.
\end{equation}
In our analysis, we fit Weibull distributions to career and collaboration lifetime data through linear regression on the corresponding cumulative distribution
\begin{equation}
F(x) = 1 - e^{-(x/\lambda)^{k}}.
\end{equation}
Taking the logarithm of both sides twice gives
\begin{equation}
\log \left( -\log (1-F(x)) \right) = k \log x - k \log \lambda
\end{equation}
This is a linear equation in $\log x$ and $\log \left( -\log (1-F(x)) \right)$. The resulting shape parameters provide insight into the underlying mechanisms of career longevity and collaboration stability in different domains.

\section{Results}
\label{sec:results}

\begin{figure*}[htb]
  \centering
  \begin{tabular}{cc}
    \includegraphics[width=0.47\textwidth]{"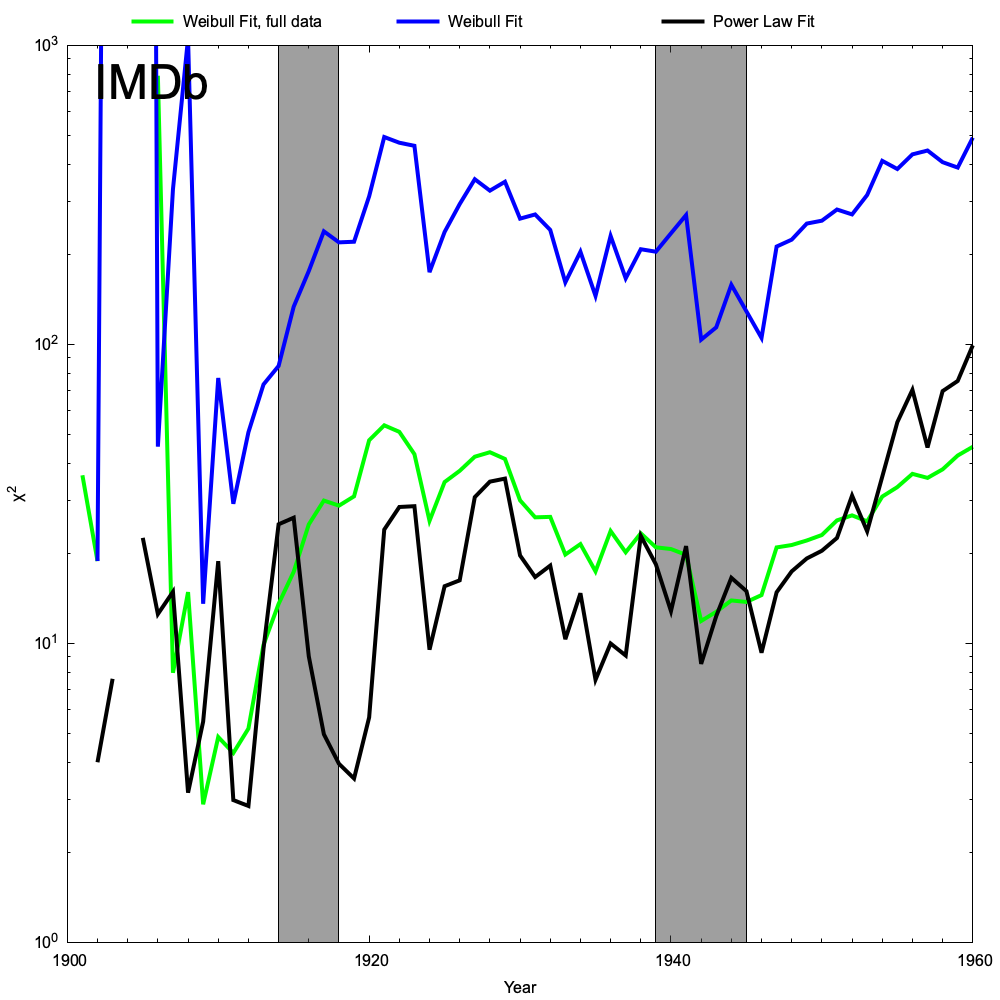"} &
    \includegraphics[width=0.47\textwidth]{"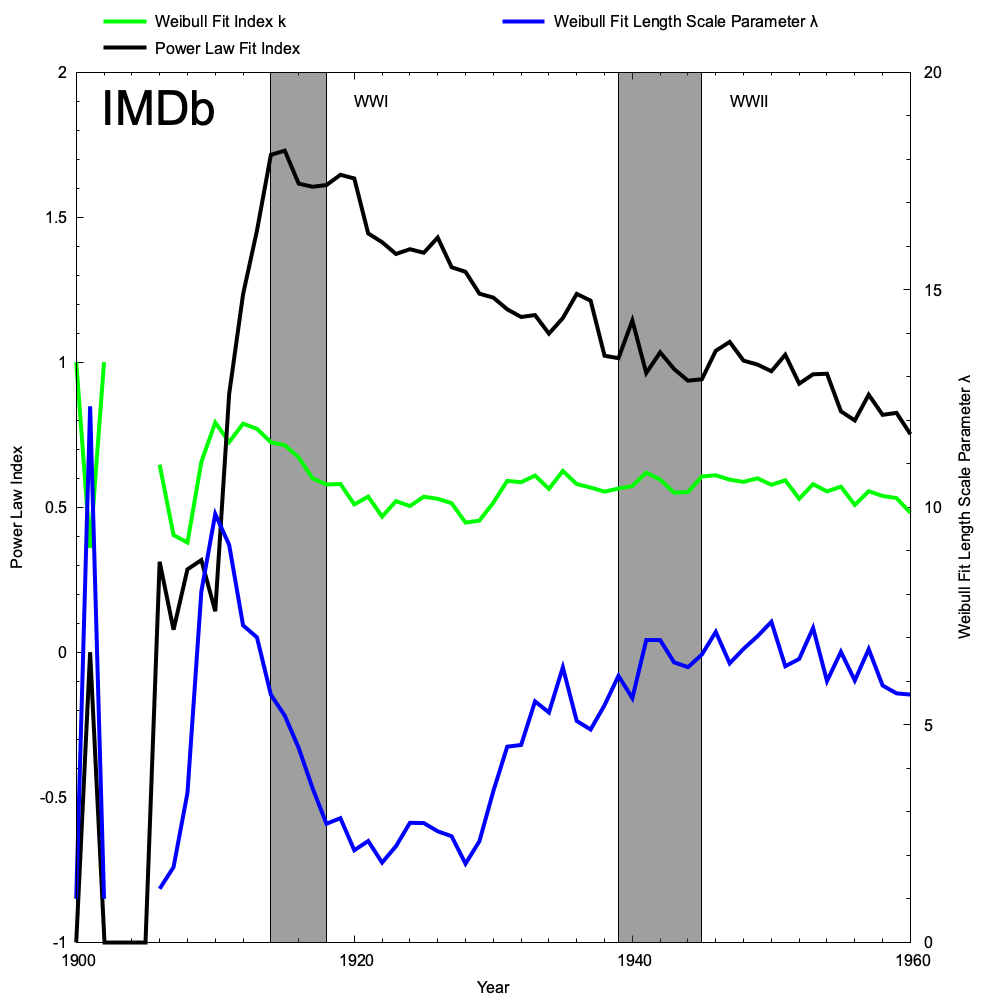"} \\
    \includegraphics[width=0.47\textwidth]{"author-career-durations-chi2.png"} &
    \includegraphics[width=0.47\textwidth]{"author-career-parameters.png"}
  \end{tabular}  
  \caption{$\chi^{2}$ and fitting parameter evolution for the career duration distributions shown in Figs. \ref{mag_careers_a} to \ref{imdb_careers_c}.}
  \label{career_fits}
\end{figure*}

\begin{figure*}[htb]
  \centering
  \begin{tabular}{cc}
    \includegraphics[width=0.47\textwidth]{"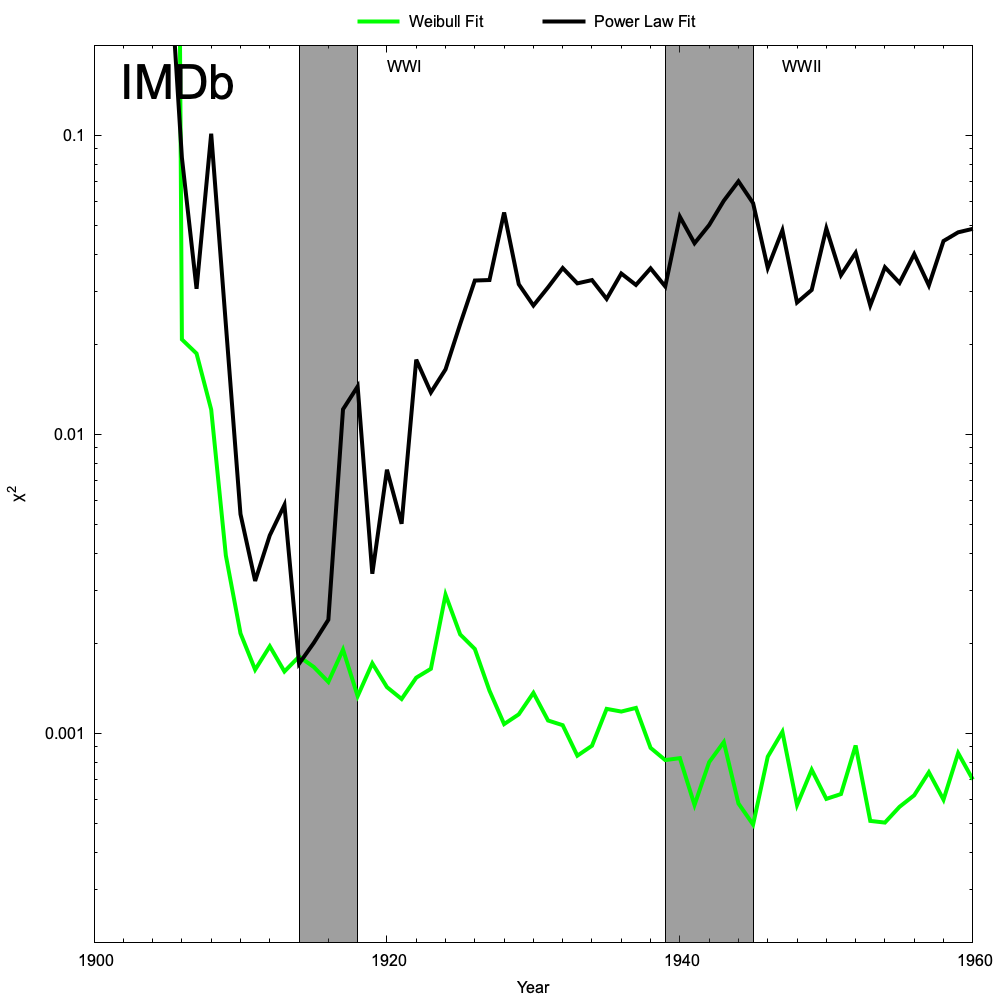"} &
    \includegraphics[width=0.47\textwidth]{"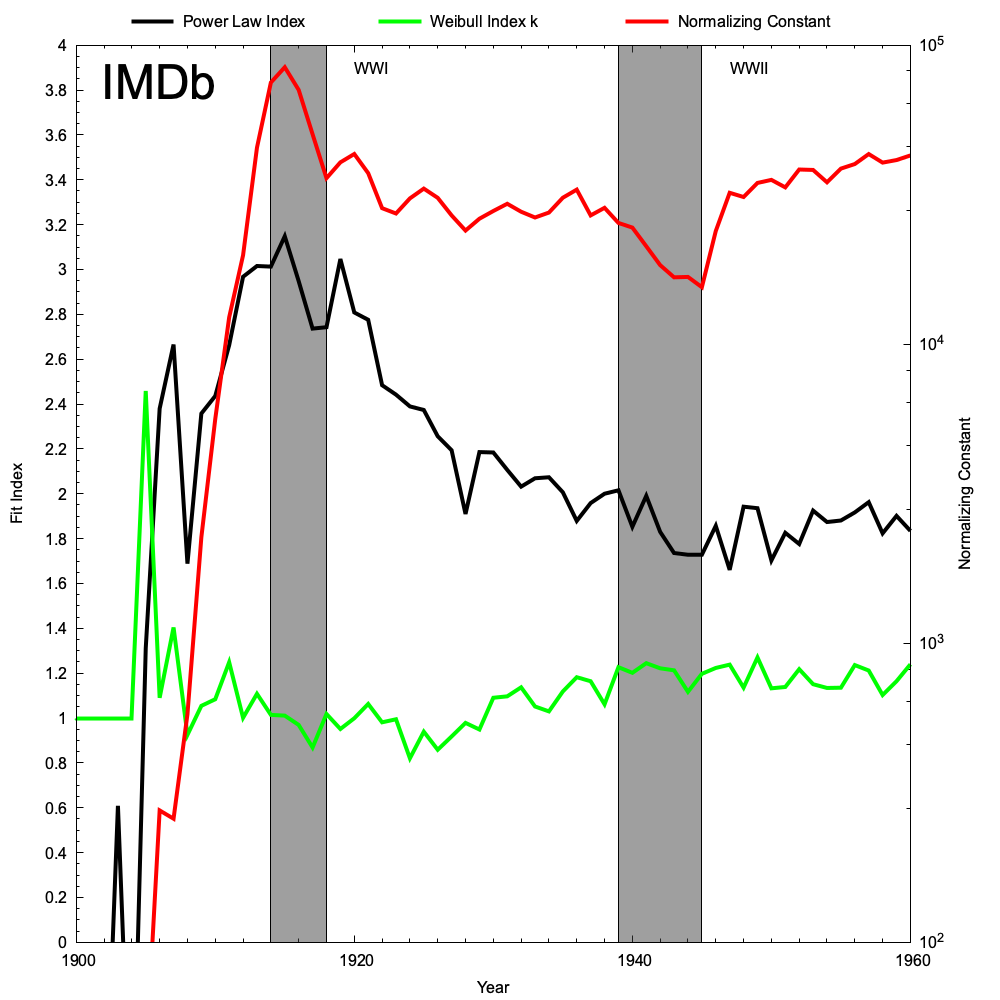"} \\
    \includegraphics[width=0.47\textwidth]{"edge-removal-chi2.png"} &
    \includegraphics[width=0.47\textwidth]{"edge-removal-parameters.png"}  
  \end{tabular}
  \caption{$\chi^{2}$ and fitting parameter evolution for the collaboration duration distributions shown in Figs. \ref{mag_edges_removal_a}-\ref{imdb_edges_removal_c}.}
  \label{edge_removal_params}
\end{figure*}

\begin{figure*}[pt]
  \centering
  \includegraphics[width=0.96\textwidth]{"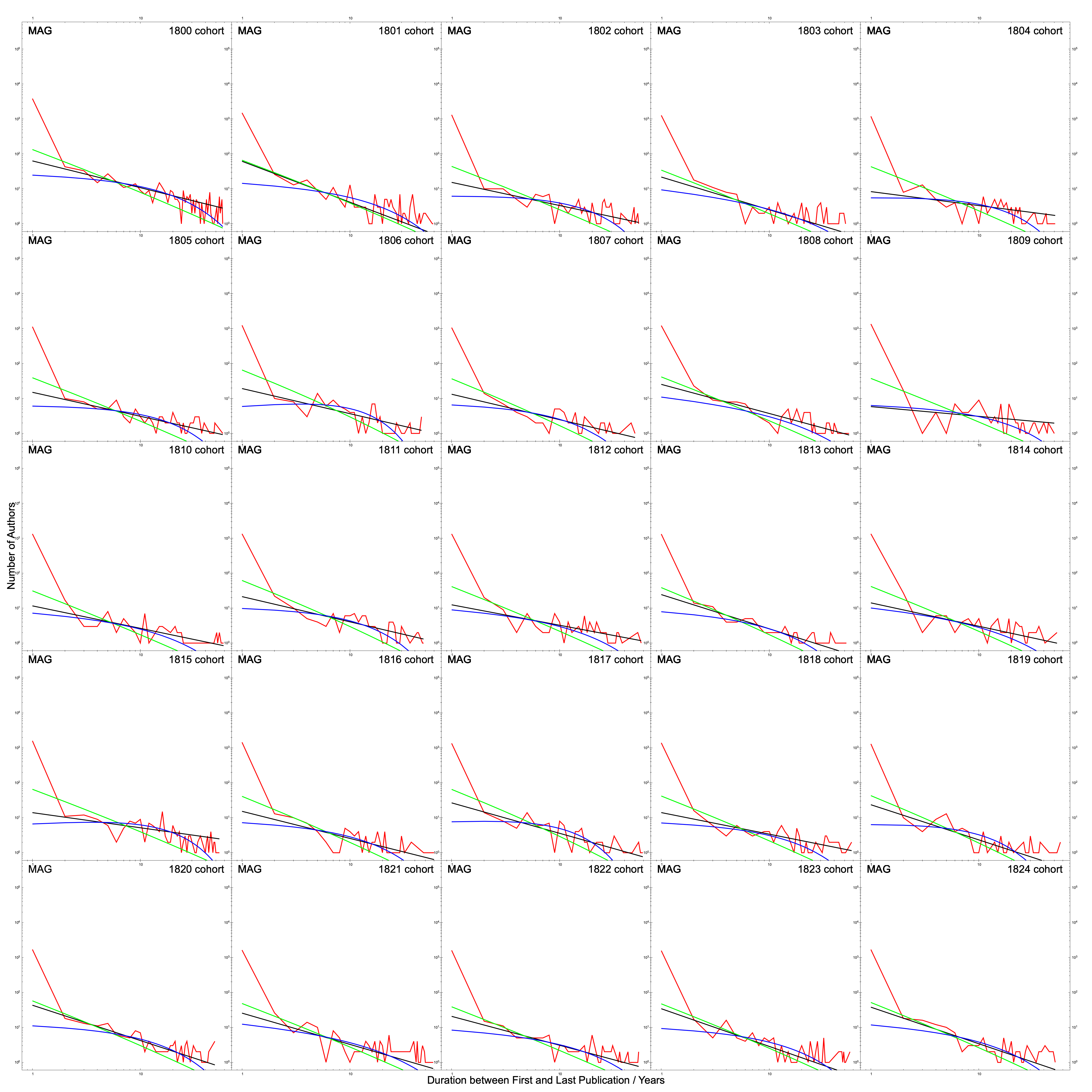"}
  \caption{Distributions of the number of authors with a given career duration, for cohorts of authors who first published in a given year, 1800 to 1824. A power-law fit, Weibull fit, and a Weibull fit excluding the central data point are shown in black, green, and blue lines respectively. Note that the y-axis scale is the same in all plots.}
  \label{mag_careers_a}
\end{figure*}

\begin{figure*}[pt]
  \centering
  \includegraphics[width=0.96\textwidth]{"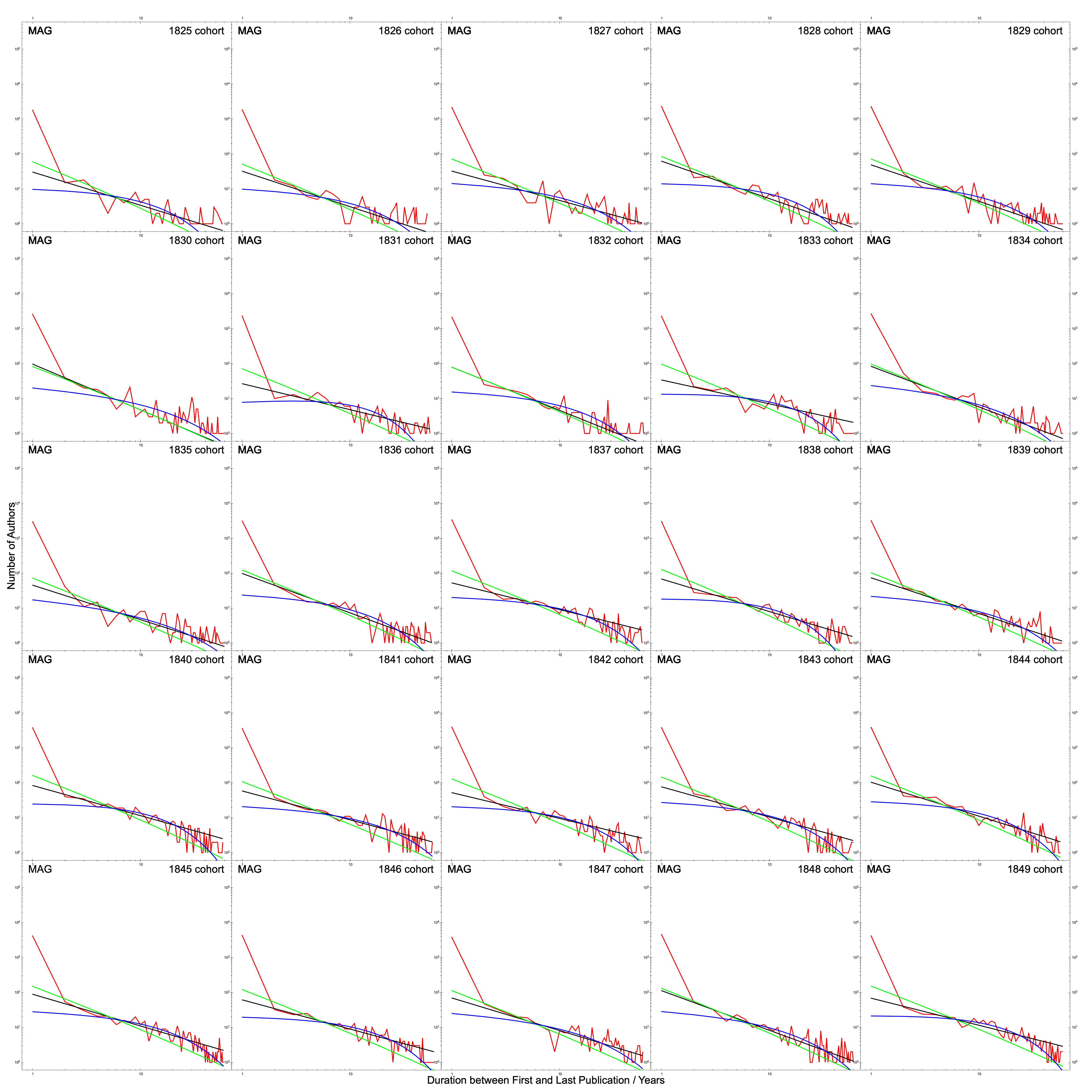"}
  \caption{Distributions of the number of authors with a given career duration, for cohorts of authors who first published in a given year, 1825 to 1859. A power-law fit, Weibull fit, and a Weibull fit excluding the central data point are shown in black, green, and blue lines respectively. Note that the y-axis scale is the same in all plots.}  
  \label{mag_careers_b}
\end{figure*}

\begin{figure*}[pt]
  \centering
  \includegraphics[width=0.96\textwidth]{"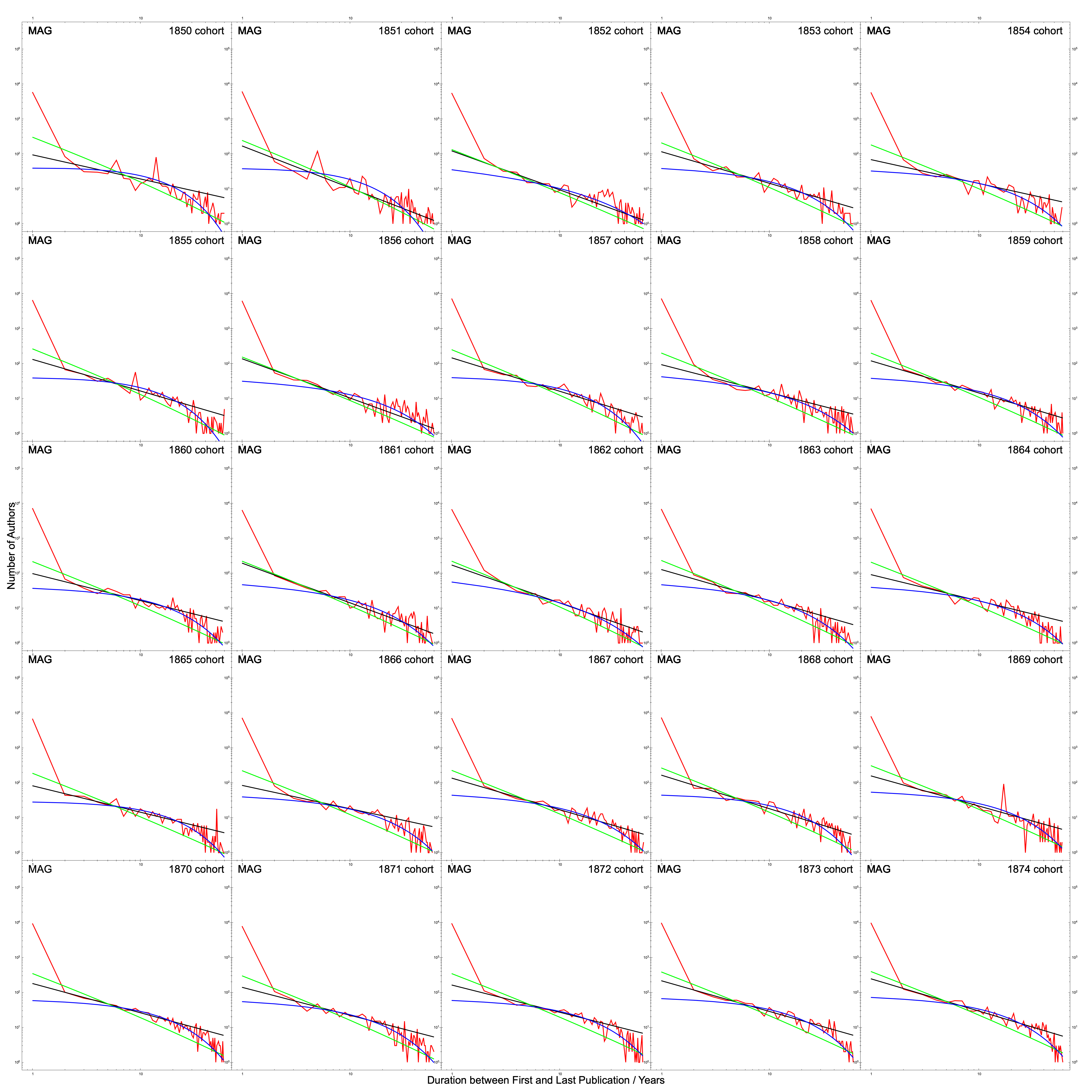"}
  \caption{Distributions of the number of authors with a given career duration, for cohorts of authors who first published in a given year, 1850 to 1874. A power-law fit, Weibull fit, and a Weibull fit excluding the central data point are shown in black, green, and blue lines respectively. Note that the y-axis scale is the same in all plots.}
  \label{mag_careers_c}
\end{figure*}

\begin{figure*}[pt]
  \centering
  \includegraphics[width=0.96\textwidth]{"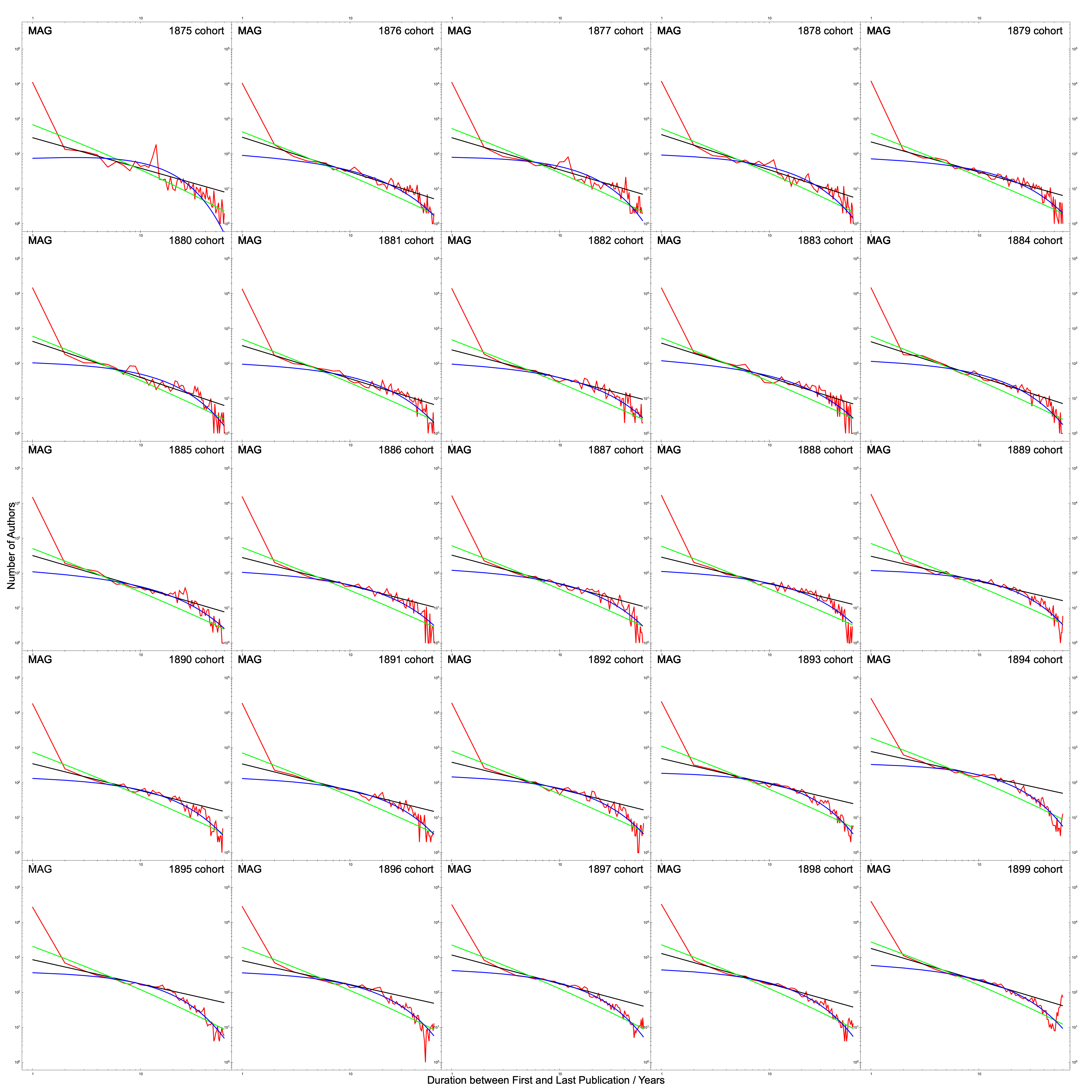"}
  \caption{Distributions of the number of authors with a given career duration, for cohorts of authors who first published in a given year, 1875 to 1899. A power-law fit, Weibull fit, and a Weibull fit excluding the central data point are shown in black, green, and blue lines respectively. Note that the y-axis scale is the same in all plots.}
  \label{mag_careers_d}
\end{figure*}

\begin{figure*}[pt]
  \centering
  \includegraphics[width=0.96\textwidth]{"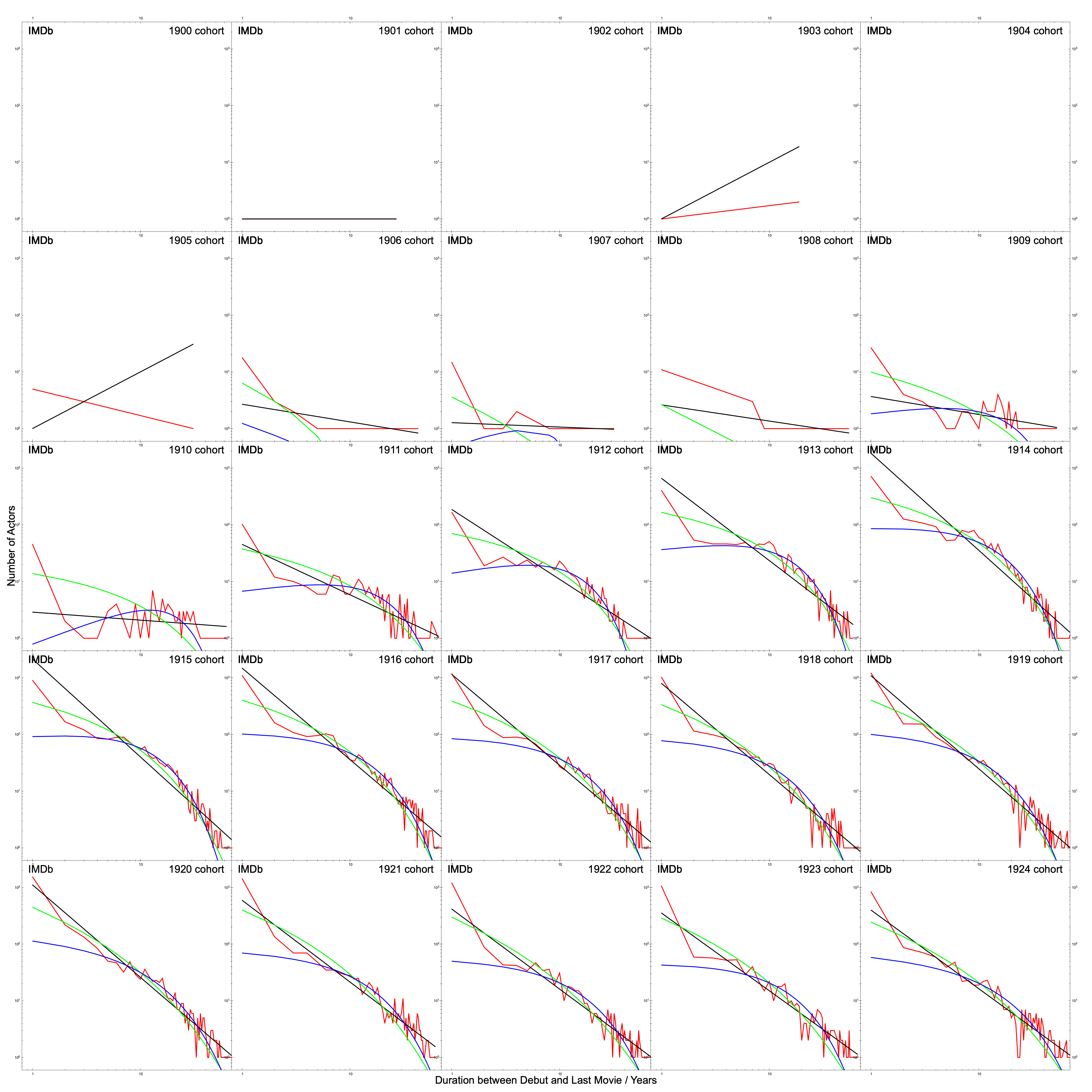"}
  \caption{Distributions of the number of authors with a given career duration, for cohorts of authors who first published in a given year, 1900 to 1924. A power-law fit, Weibull fit, and a Weibull fit excluding the central data point are shown in black, green, and blue lines respectively. Note that the y-axis scale is the same in all plots.}
  \label{mag_careers_e}
\end{figure*}

\begin{figure*}[pt]
  \centering
  \includegraphics[width=0.96\textwidth]{"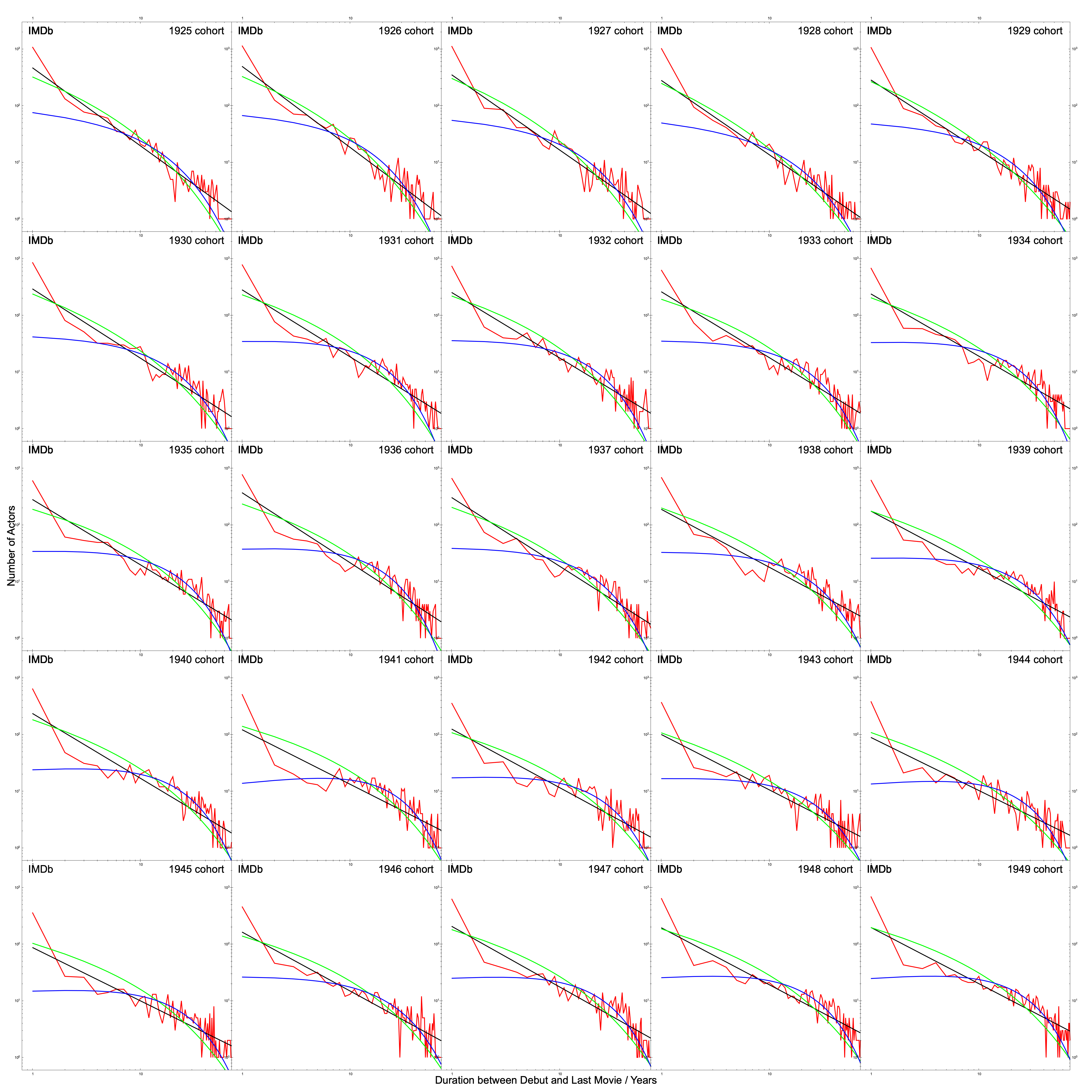"}
  \caption{Distributions of the number of authors with a given career duration, for cohorts of authors who first published in a given year, 1925 to 1949. A power-law fit, Weibull fit, and a Weibull fit excluding the central data point are shown in black, green, and blue lines respectively. Note that the y-axis scale is the same in all plots.}
  \label{mag_careers_f}
\end{figure*}

\begin{figure*}[pt]
  \centering
  \includegraphics[width=0.96\textwidth]{"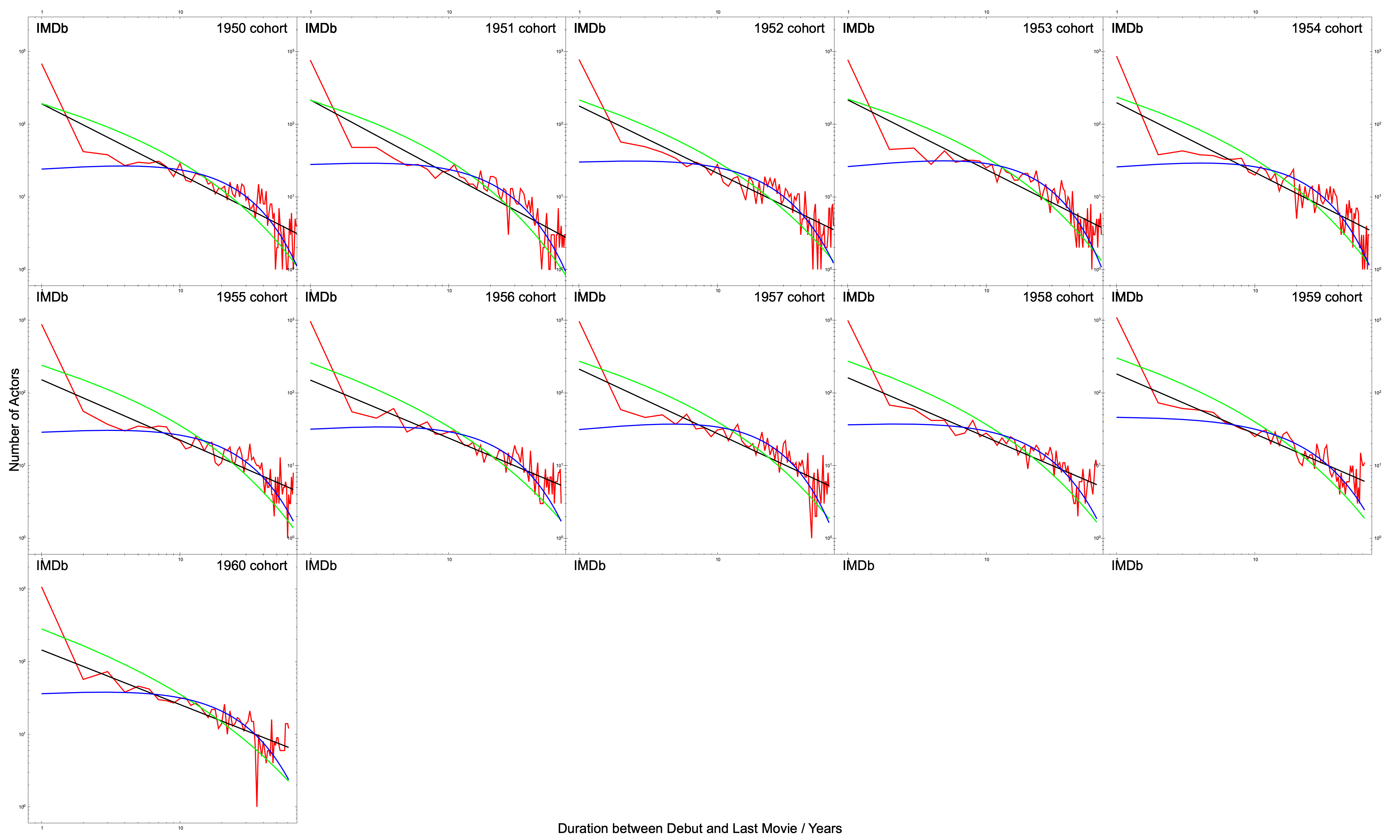"}
  \caption{Distributions of the number of authors with a given career duration, for cohorts of authors who first published in a given year, 1950 to 1964. A power-law fit, Weibull fit, and a Weibull fit excluding the central data point are shown in black, green, and blue lines respectively. Note that the y-axis scale is the same in all plots.}
  \label{mag_careers_g}
\end{figure*}

\begin{figure*}[pt]
  \centering
  \includegraphics[width=0.96\textwidth]{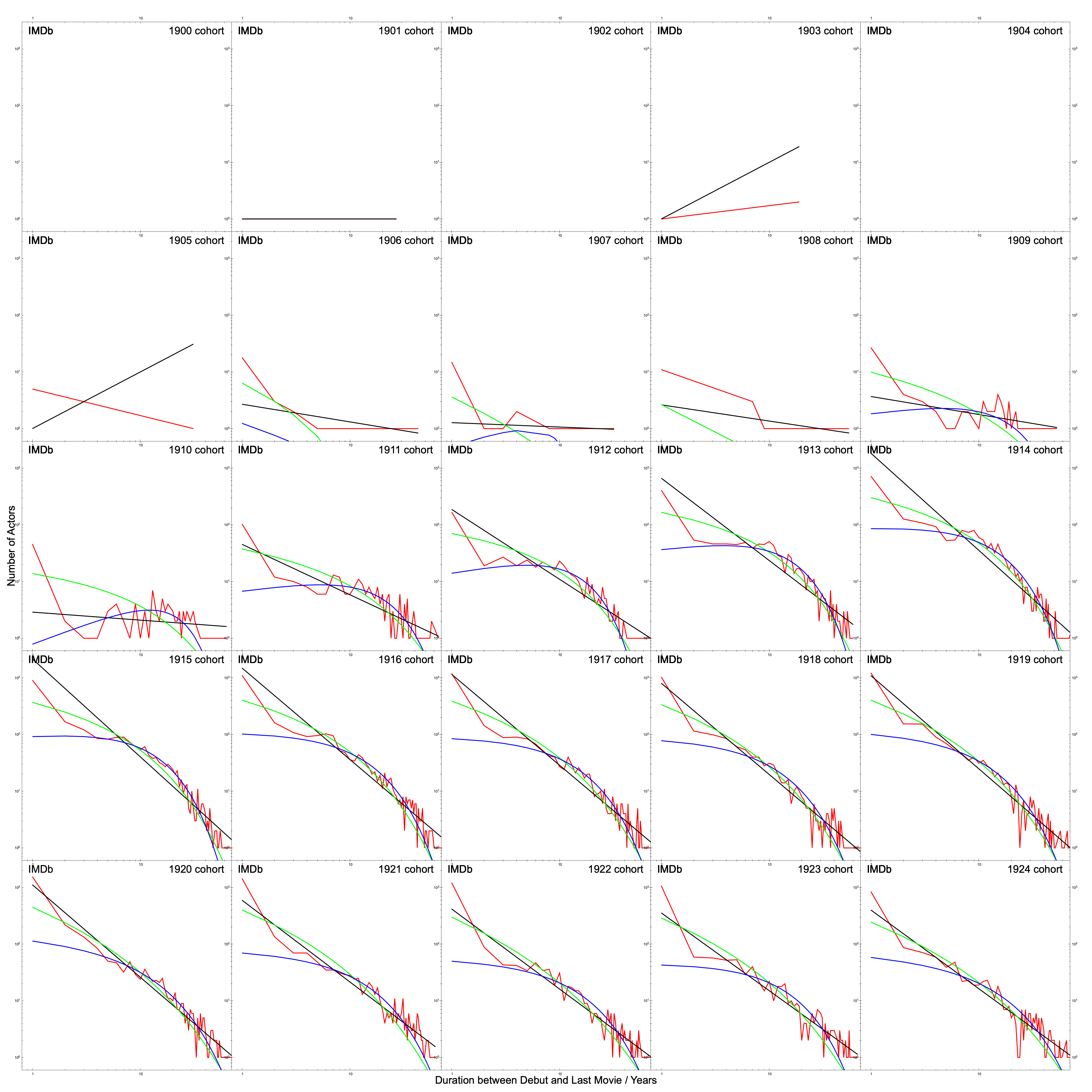}
  \caption{Distributions of the number of authors with a given career duration, for cohorts of authors who first published in a given year, 1900 to 1924. A power-law fit, Weibull fit, and a Weibull fit excluding the central data point are shown in black, green, and blue lines respectively. Note that the y-axis scale is the same in all plots.}
  \label{imdb_careers_a}
\end{figure*}

\begin{figure*}[pt]
  \centering
  \includegraphics[width=0.96\textwidth]{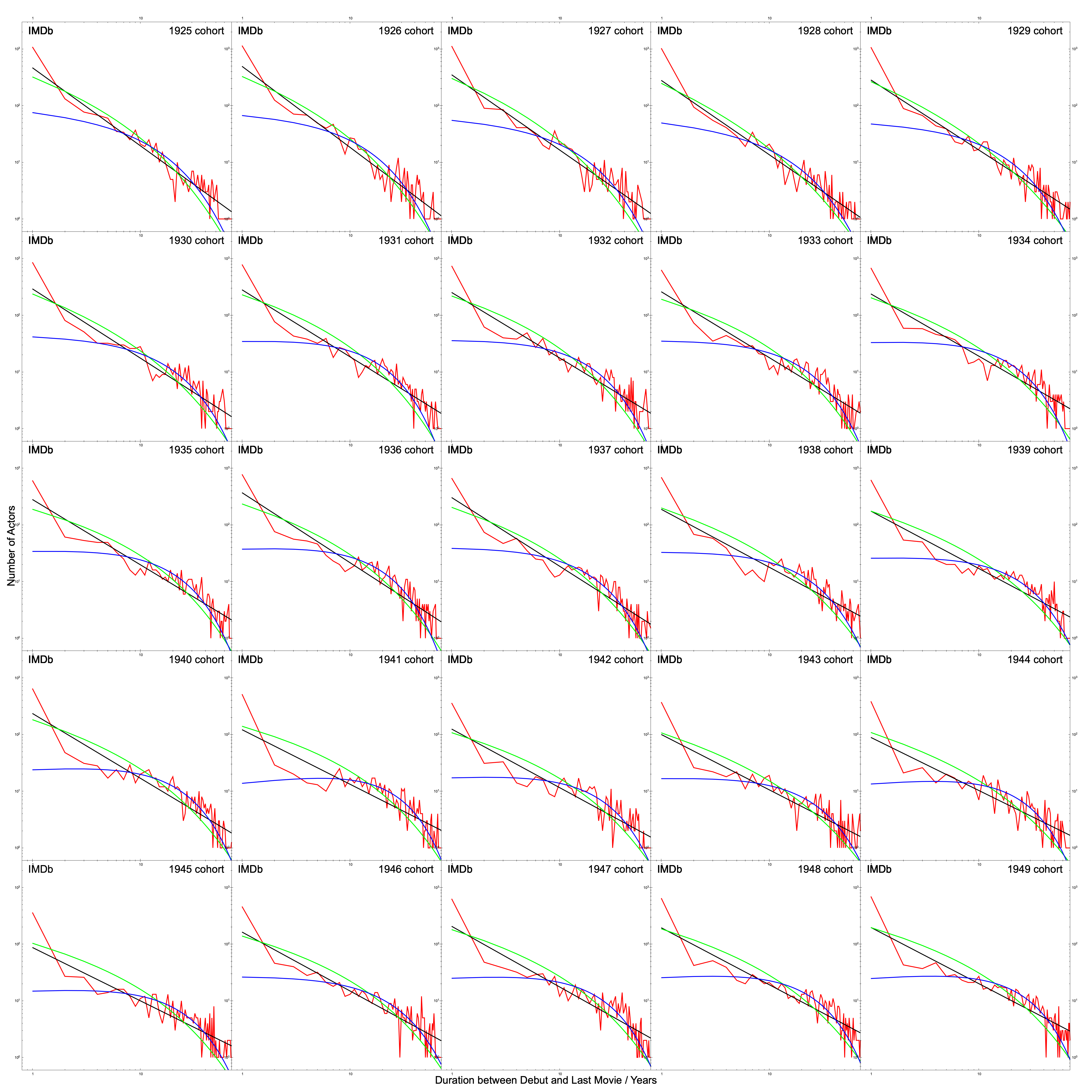}
  \caption{Distributions of the number of authors with a given career duration, for cohorts of authors who first published in a given year, 1925 to 1949. A power-law fit, Weibull fit, and a Weibull fit excluding the central data point are shown in black, green, and blue lines respectively. Note that the y-axis scale is the same in all plots.}
  \label{imdb_careers_b}
\end{figure*}

\begin{figure*}[pt]
  \centering
  \includegraphics[width=0.96\textwidth]{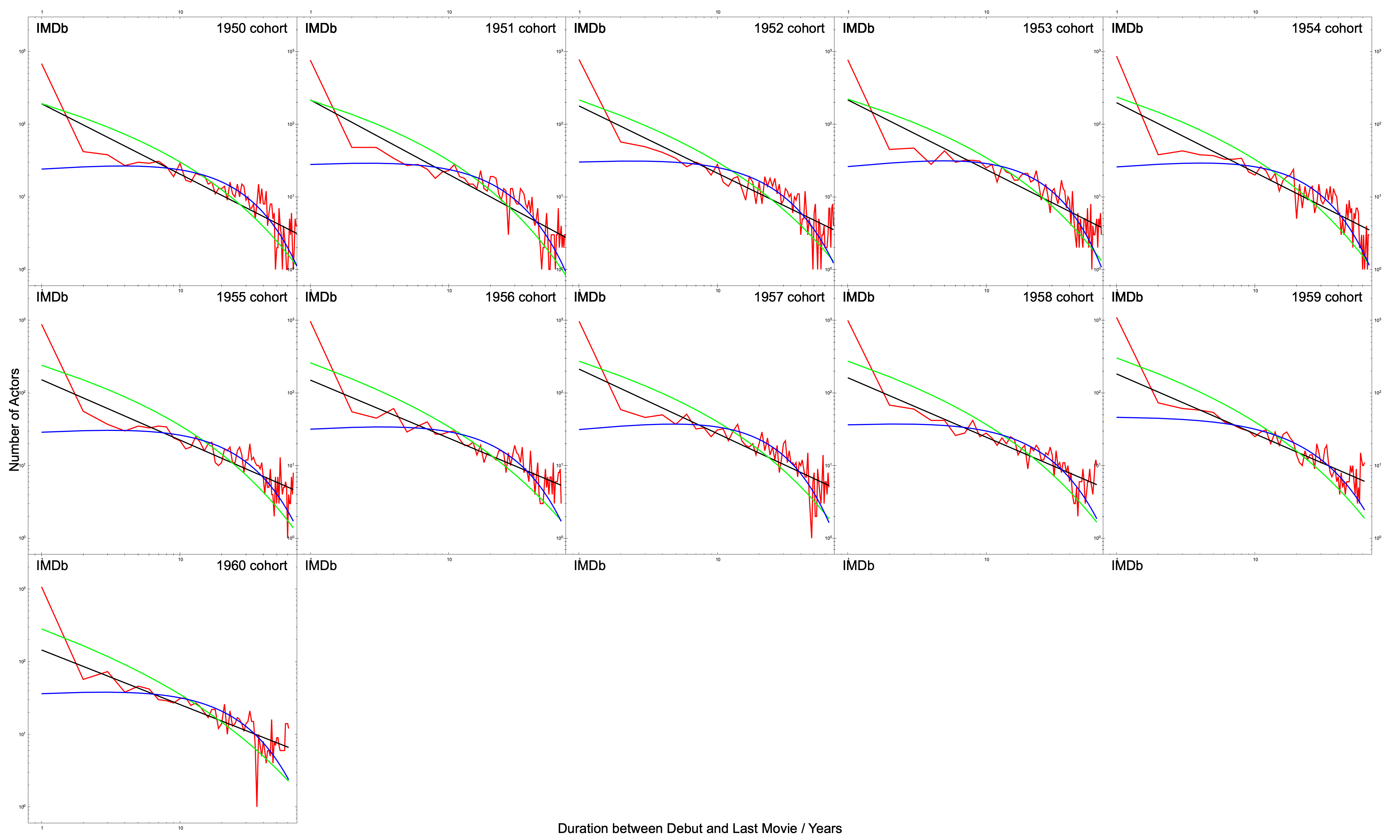}
  \caption{Distributions of the number of authors with a given career duration, for cohorts of authors who first published in a given year, 1950 to 1964. A power-law fit, Weibull fit, and a Weibull fit excluding the central data point are shown in black, green, and blue lines respectively. Note that the y-axis scale is the same in all plots.}
  \label{imdb_careers_c}
\end{figure*}

\begin{figure*}[pt]
  \centering
  \begin{tabular}{c}
    \includegraphics[width=0.96\textwidth]{"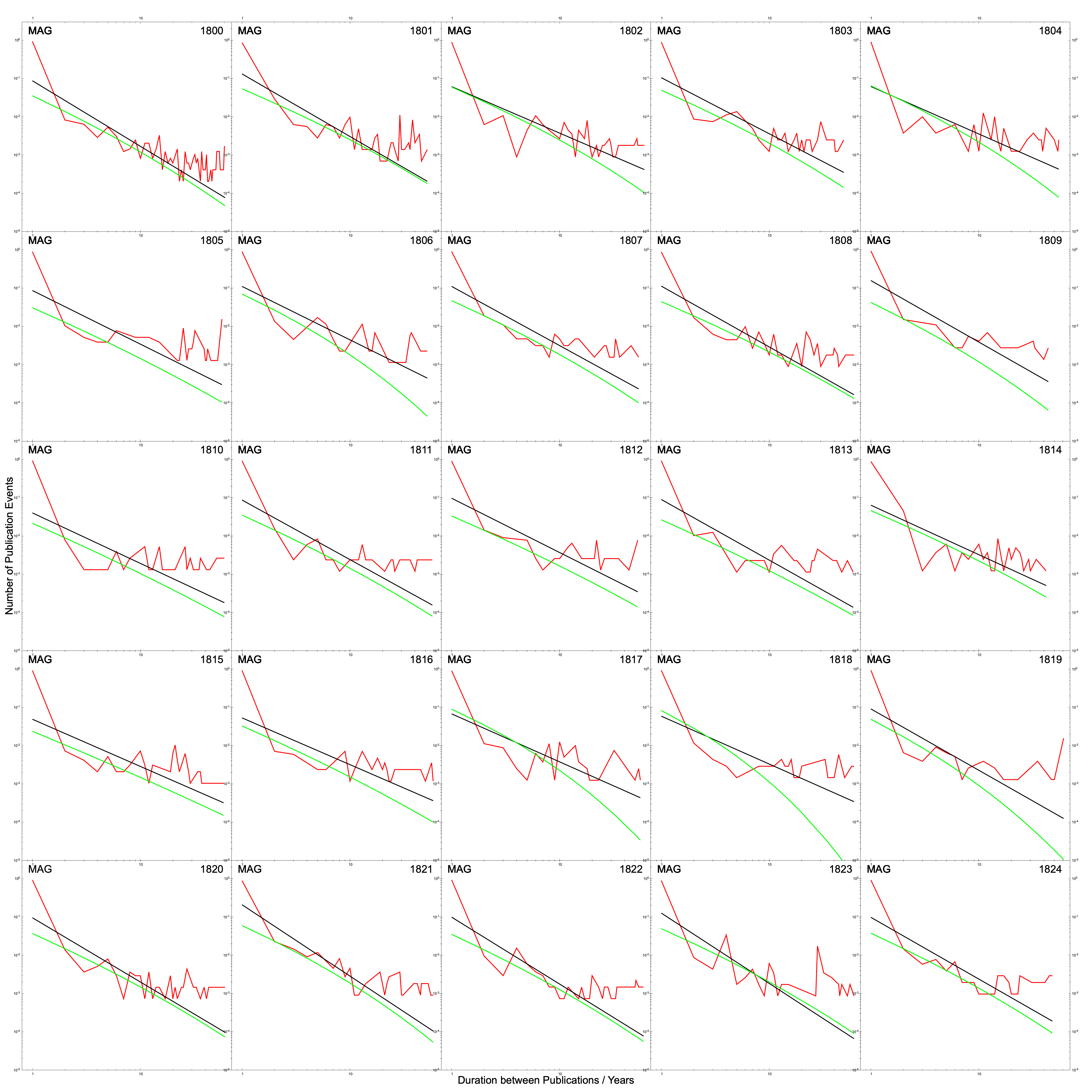"}
  \end{tabular}
  \caption{Distributions of collaboration duration, for cohorts of authors who first published in a given year, 1800 to 1824. A power-law fit, Weibull fit, and a Weibull fit excluding the central data point are shown in black, green, and blue lines respectively. Note that the y-axis scale is the same in all plots.}
  \label{mag_edges_removal_a}
\end{figure*}

\begin{figure*}[pt]
  \centering
  \begin{tabular}{c}
    \includegraphics[width=0.96\textwidth]{"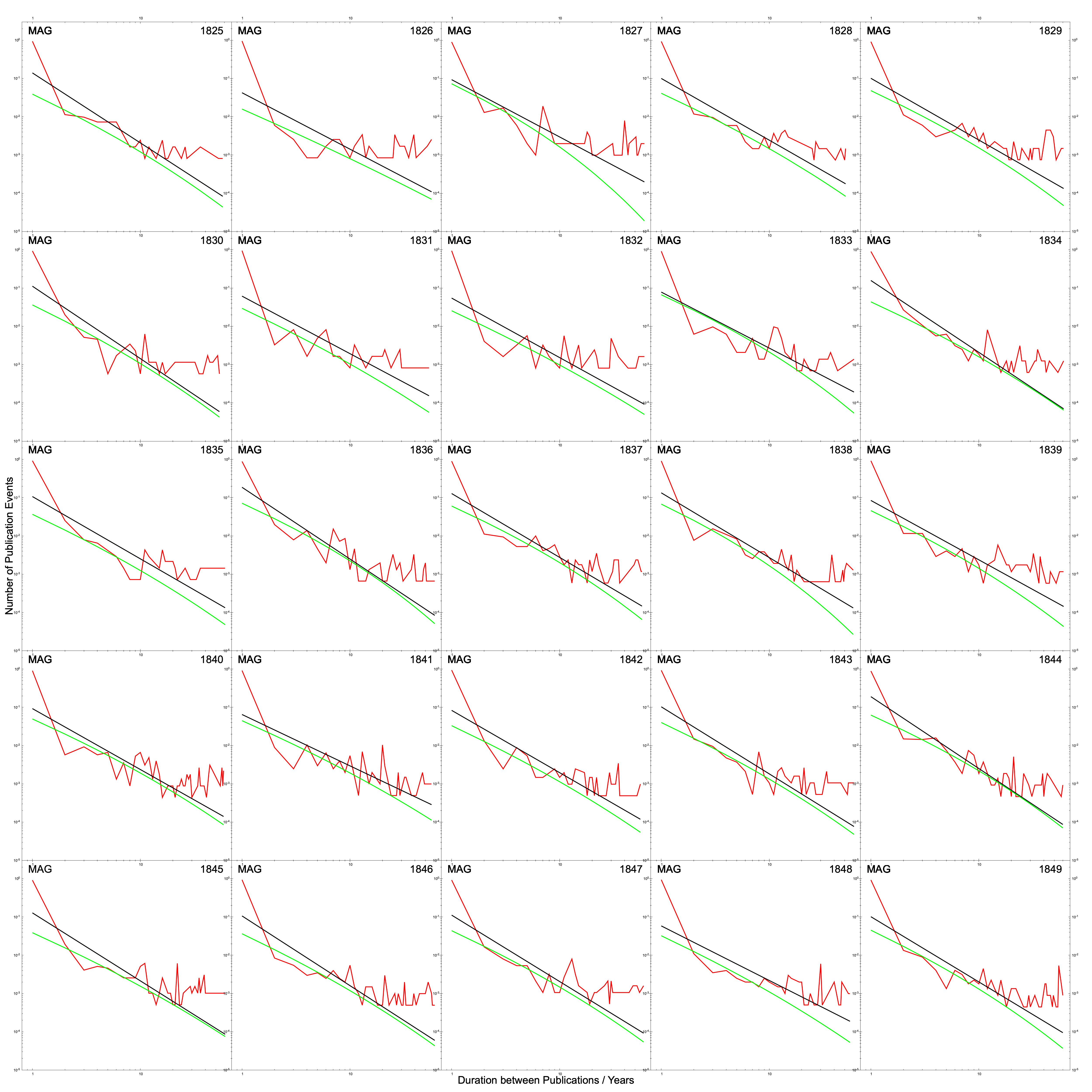"}
  \end{tabular}
  \caption{Distributions of collaboration duration, for cohorts of authors who first published in a given year, 1825 to 1849. A power-law fit, Weibull fit, and a Weibull fit excluding the central data point are shown in black, green, and blue lines respectively. Note that the y-axis scale is the same in all plots.}
  \label{mag_edges_removal_b}
\end{figure*}

\begin{figure*}[pt]
  \centering
  \begin{tabular}{c}
    \includegraphics[width=0.96\textwidth]{"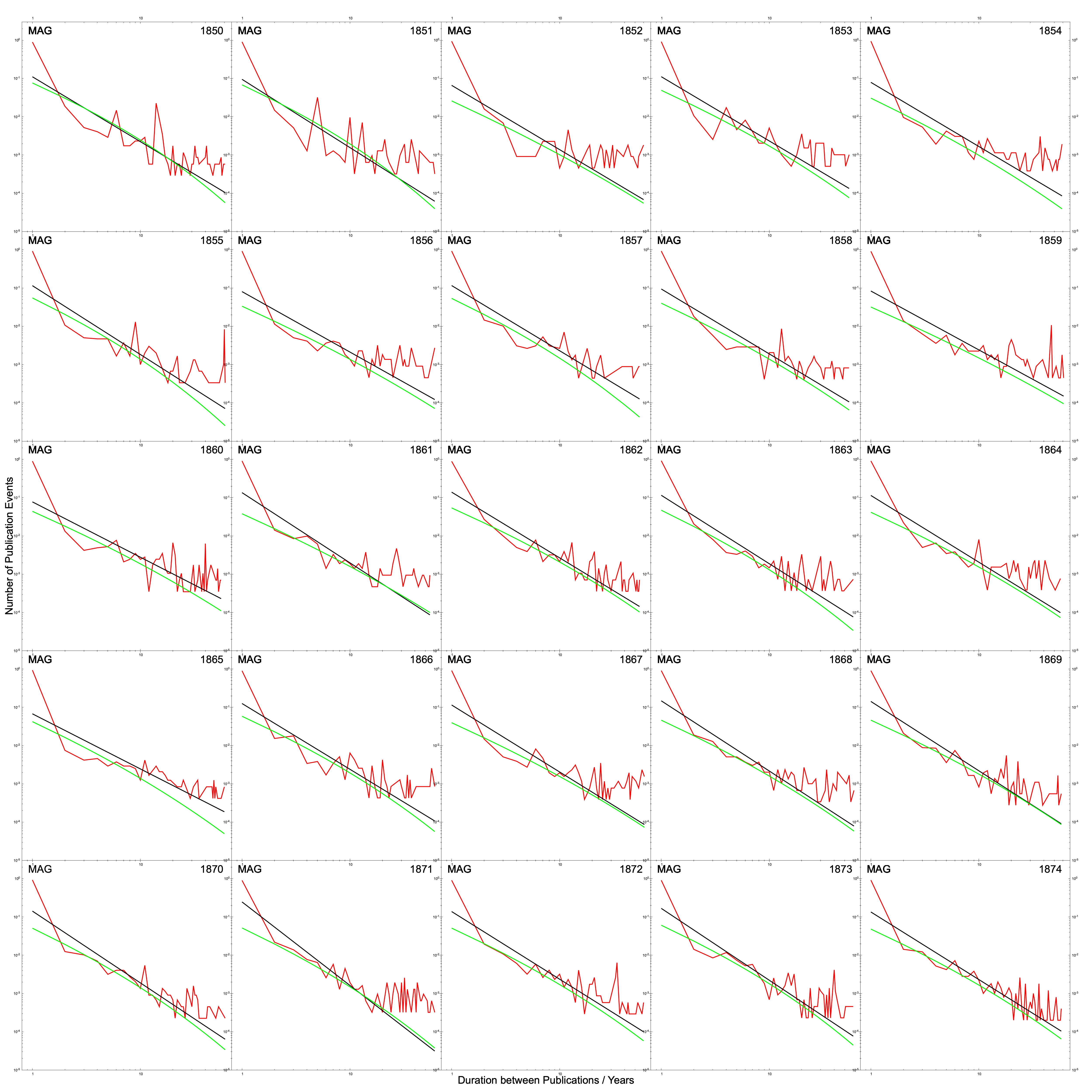"}
  \end{tabular}
  \caption{Distributions of collaboration duration, for cohorts of authors who first published in a given year, 1850 to 1874. A power-law fit, Weibull fit, and a Weibull fit excluding the central data point are shown in black, green, and blue lines respectively. Note that the y-axis scale is the same in all plots.}  
  \label{mag_edges_removal_c}
\end{figure*}

\begin{figure*}[pt]
  \centering
  \begin{tabular}{c}
    \includegraphics[width=0.96\textwidth]{"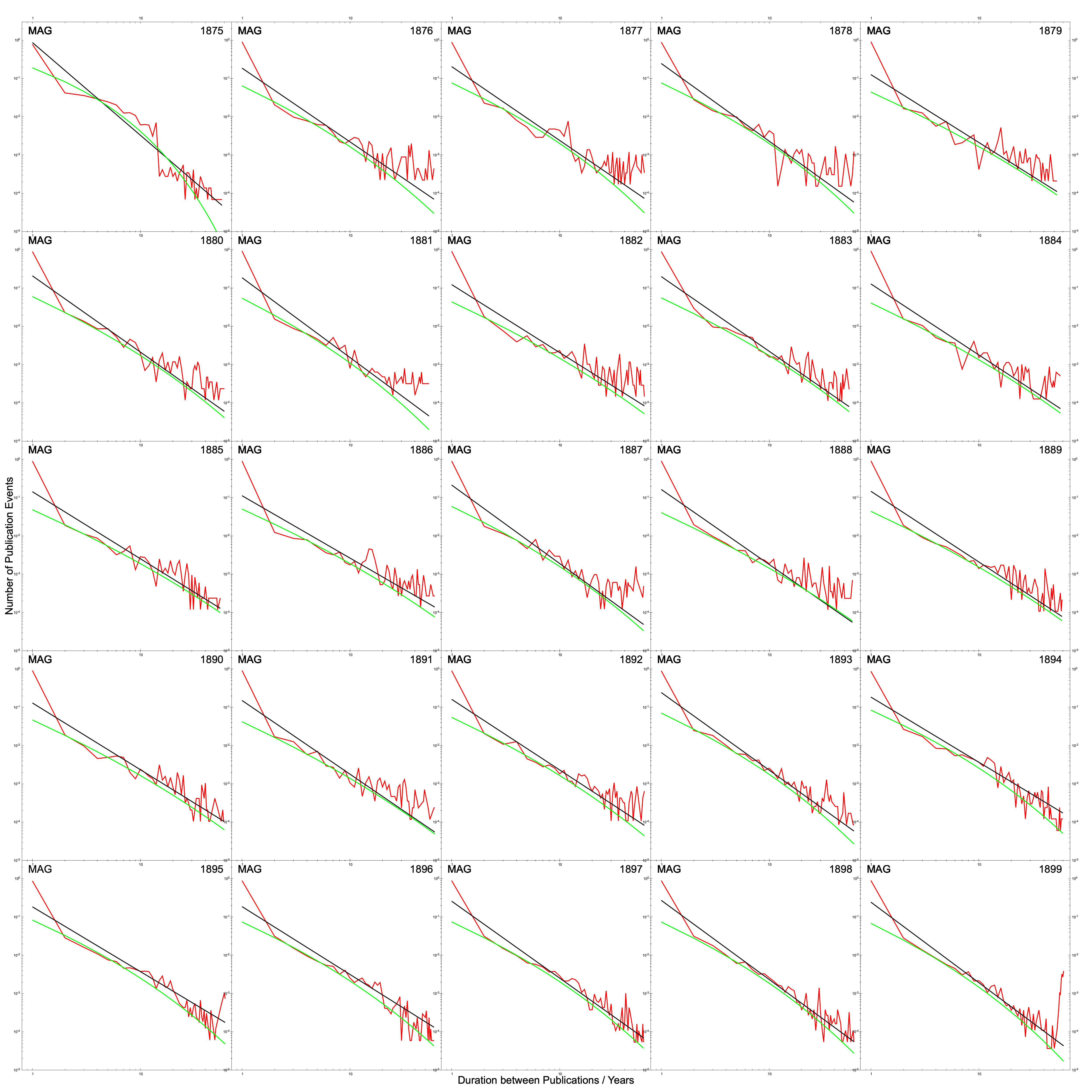"}
  \end{tabular}
  \caption{Distributions of collaboration duration, for cohorts of authors who first published in a given year, 1875 to 1899. A power-law fit, Weibull fit, and a Weibull fit excluding the central data point are shown in black, green, and blue lines respectively. Note that the y-axis scale is the same in all plots.}
  \label{mag_edges_removal_d}
\end{figure*}

\begin{figure*}[pt]
  \centering
  \begin{tabular}{c}
    \includegraphics[width=0.96\textwidth]{"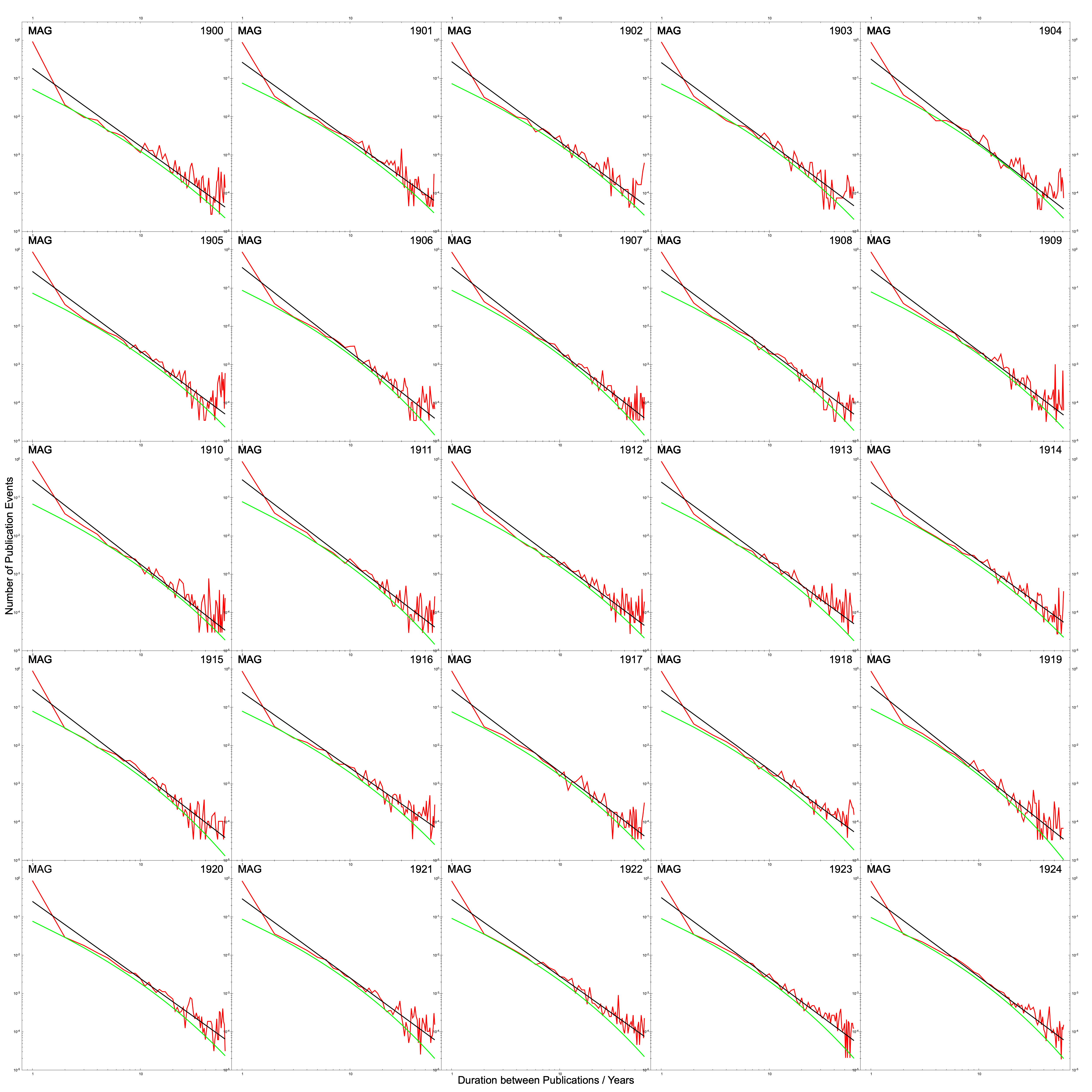"}
  \end{tabular}
  \caption{Distributions of collaboration duration, for cohorts of authors who first published in a given year, 1900 to 1924. A power-law fit, Weibull fit, and a Weibull fit excluding the central data point are shown in black, green, and blue lines respectively. Note that the y-axis scale is the same in all plots.}
  \label{mag_edges_removal_e}
\end{figure*}

\begin{figure*}[pt]
  \centering
  \begin{tabular}{c}
    \includegraphics[width=0.96\textwidth]{"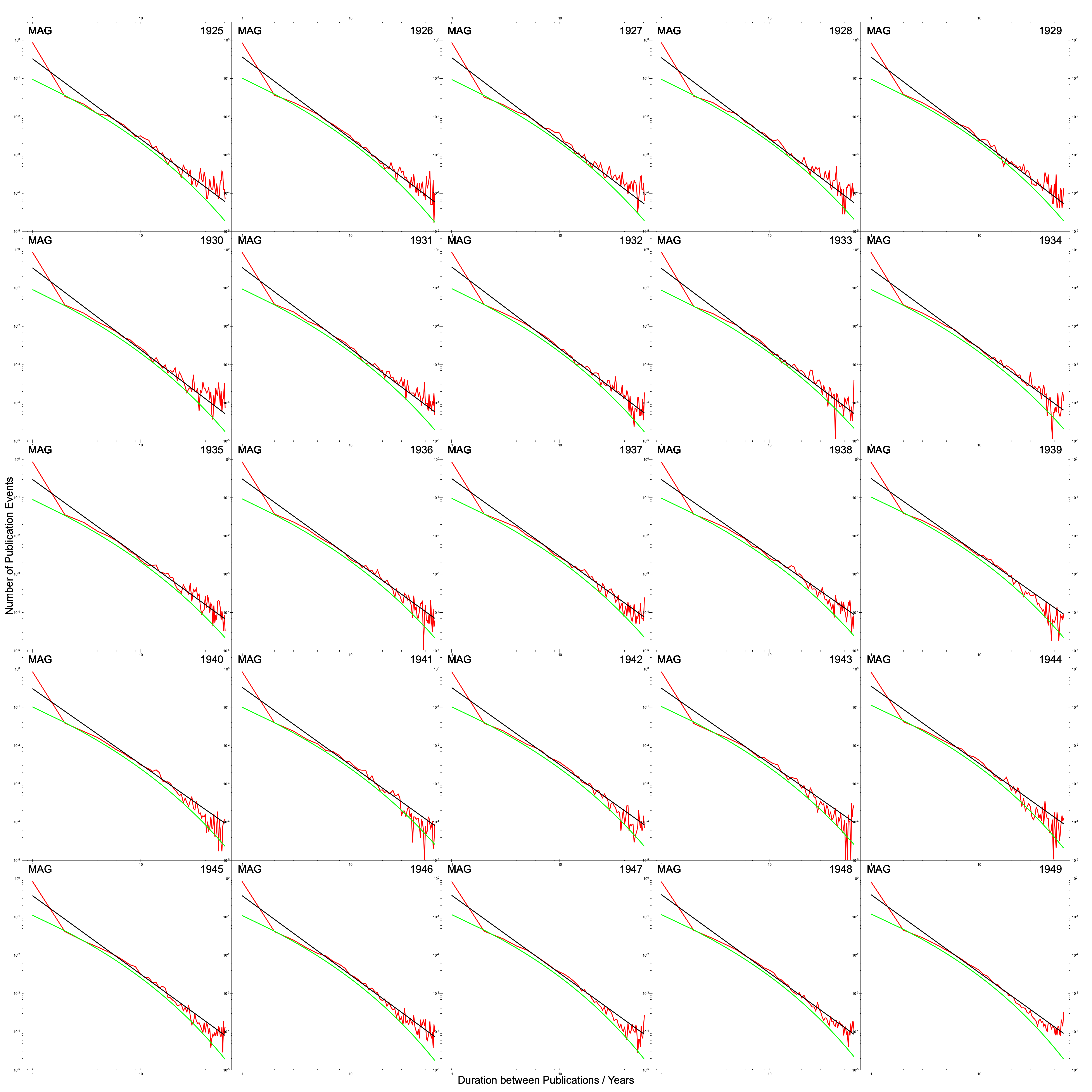"}
  \end{tabular}
  \caption{Distributions of collaboration duration, for cohorts of authors who first published in a given year, 1925 to 1949. A power-law fit, Weibull fit, and a Weibull fit excluding the central data point are shown in black, green, and blue lines respectively. Note that the y-axis scale is the same in all plots.}
  \label{mag_edges_removal_f}
\end{figure*}

\begin{figure*}[pt]
  \centering
  \begin{tabular}{c}
    \includegraphics[width=0.96\textwidth]{"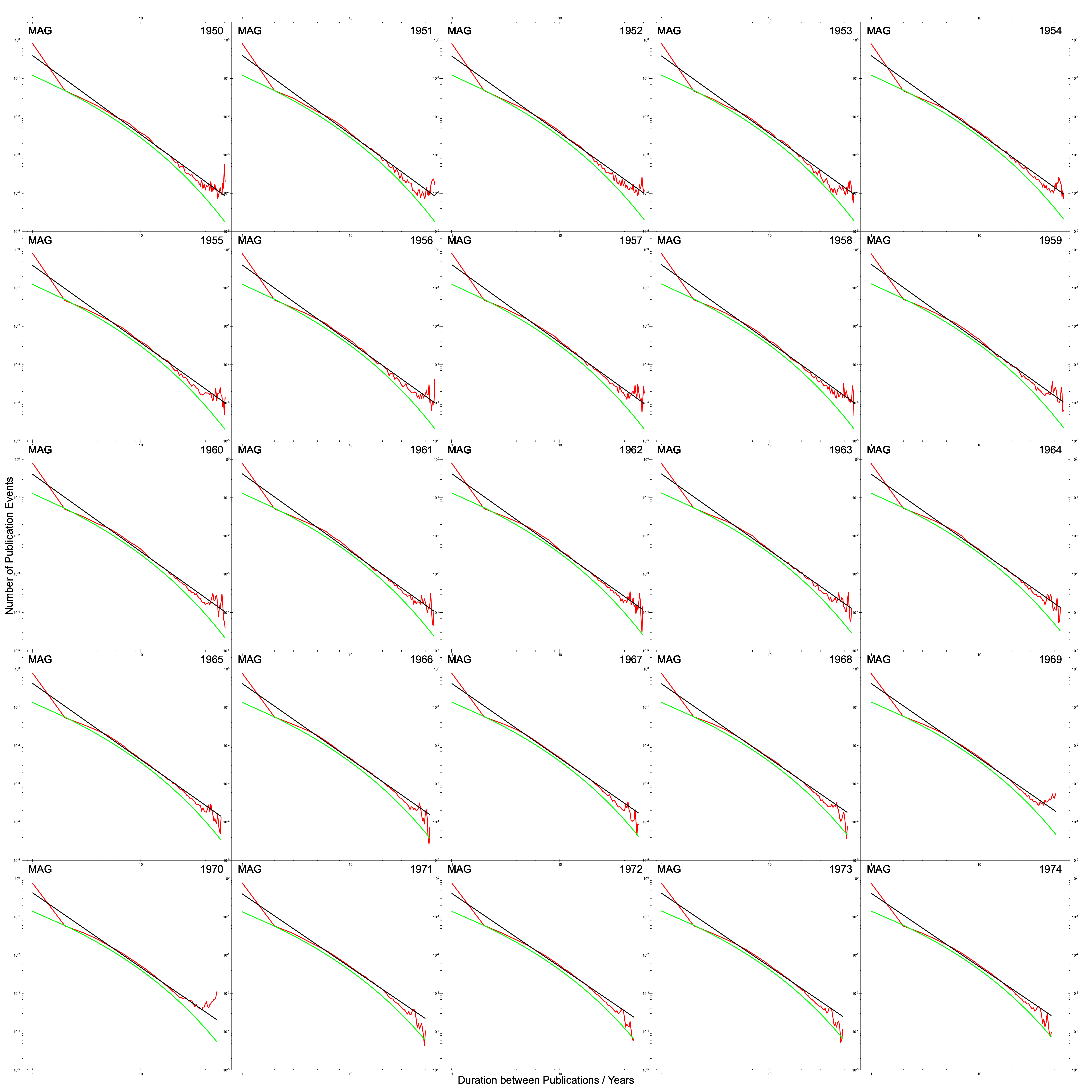"}
  \end{tabular}
  \caption{Distributions of collaboration duration, for cohorts of authors who first published in a given year, 1950 to 1974. A power-law fit, Weibull fit, and a Weibull fit excluding the central data point are shown in black, green, and blue lines respectively. Note that the y-axis scale is the same in all plots.}
  \label{mag_edges_removal_g}
\end{figure*}

\begin{figure*}[pt]
  \centering
  \begin{tabular}{c}
    \includegraphics[width=0.96\textwidth]{"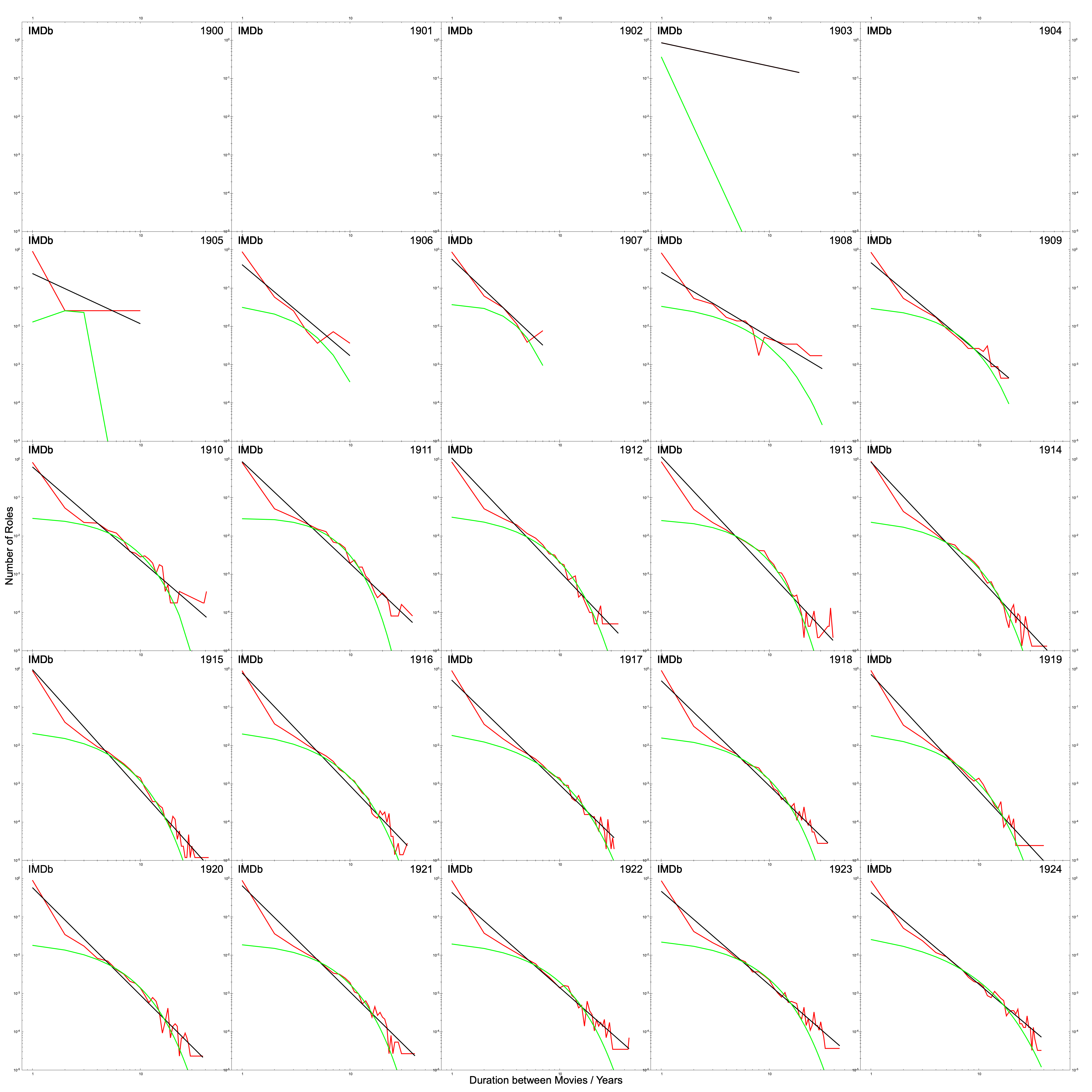"}
  \end{tabular}
  \caption{Distributions of collaboration duration, for cohorts of lead actors who first starred in a movie in a given year, 1900 to 1924. AA power-law fit, Weibull fit, and a Weibull fit excluding the central data point are shown in black, green, and blue lines respectively. Note that the y-axis scale is the same in all plots.}
  \label{imdb_edges_removal_a}
\end{figure*}

\begin{figure*}[pt]
  \centering
  \begin{tabular}{c}
    \includegraphics[width=0.96\textwidth]{"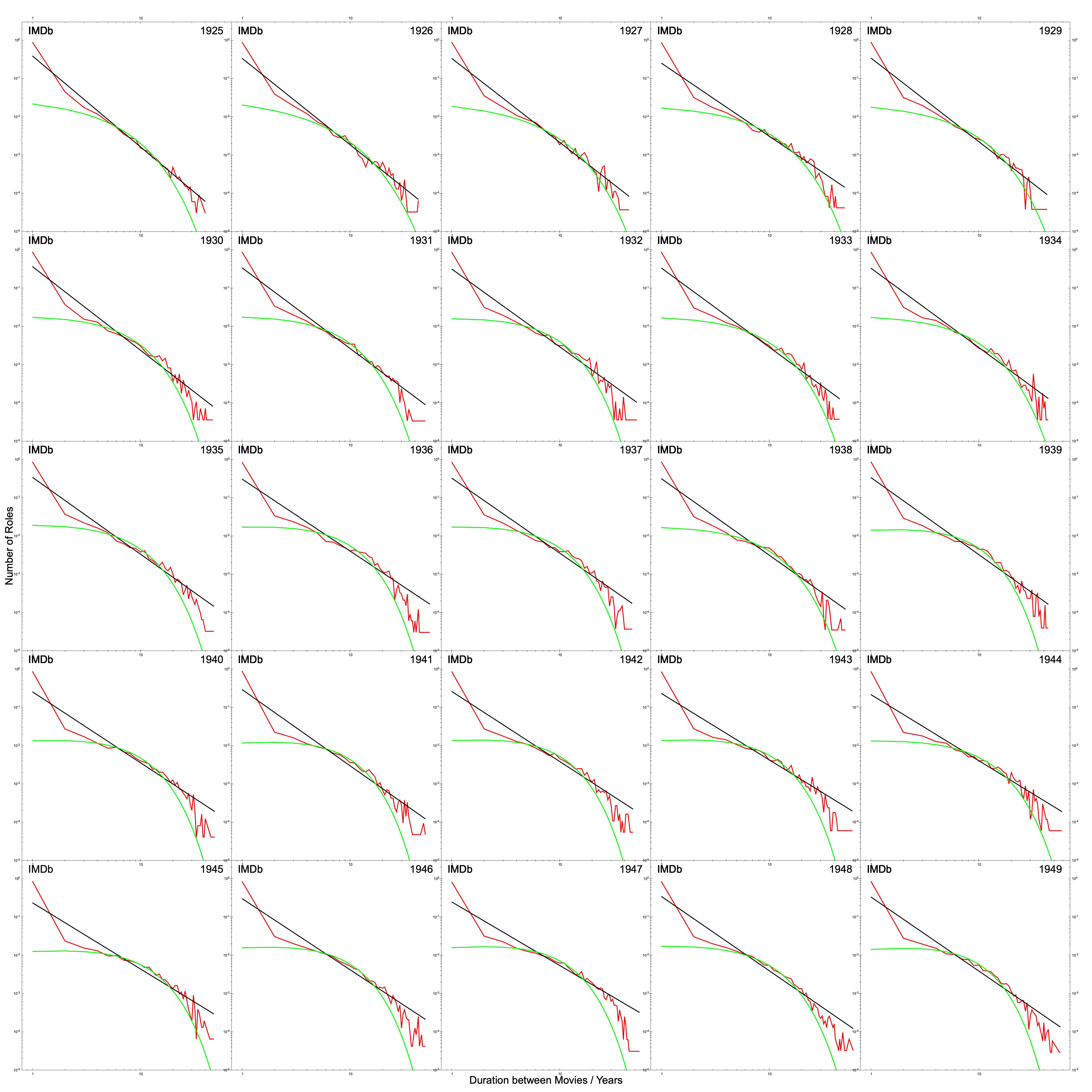"}
  \end{tabular}
  \caption{Distributions of collaboration duration, for cohorts of lead actors who first starred in a movie in a given year, 1925 to 1949. A power-law fit, Weibull fit, and a Weibull fit excluding the central data point are shown in black, green, and blue lines respectively. Note that the y-axis scale is the same in all plots.}
  \label{imdb_edges_removal_b}
\end{figure*}

\begin{figure*}[pt]
  \centering
  \begin{tabular}{c}
    \includegraphics[width=0.96\textwidth]{"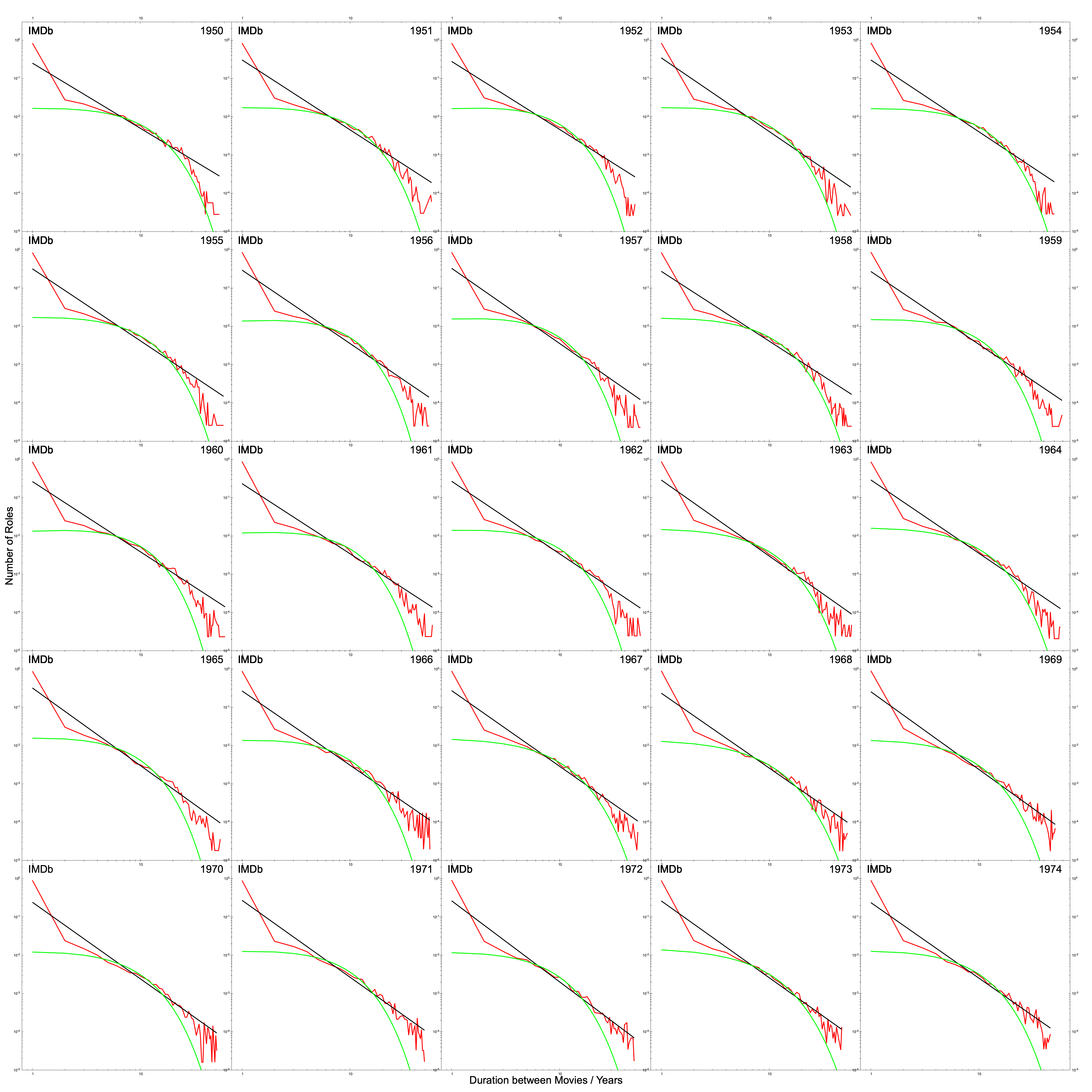"}
  \end{tabular}
  \caption{Distributions of collaboration duration, for cohorts of lead actors who first starred in a movie in a given year, 1950 to 1974. A power-law fit, Weibull fit, and a Weibull fit excluding the central data point are shown in black, green, and blue lines respectively. Note that the y-axis scale is the same in all plots.}
  \label{imdb_edges_removal_c}
\end{figure*}

Our analysis reveals distinct patterns in how careers and collaborations evolve across academic and entertainment networks. The complete distributions of lifetimes for both nodes and edges are presented in Figures \ref{mag_careers_a} to \ref{imdb_careers_c} and \ref{mag_edges_removal_a} to \ref{imdb_edges_removal_c}. For each distribution, we fitted both power-law and Weibull functions, with their parameter evolution shown in Figures \ref{career_fits} and \ref{edge_removal_params}.

\subsection{Evolution of Academic Careers and Participation Patterns}

The Microsoft Academic Graph (MAG) reveals a complex picture of how academic careers have evolved over two centuries. While the absolute number of single-year authors has grown exponentially—from roughly 100 per year in 1800 to over $10^5$ by 2020—the relative proportion of these transient participants has decreased. The fraction of single-year authors declined from approximately 0.8 in early periods to 0.6 in recent years, suggesting that modern researchers are more likely to continue their academic careers beyond their initial publication.

Career duration distributions in academia show three distinct historical phases:

\subsubsection{Early Period (1800-1874)}
This period exhibits noisy distributions because of smaller sample sizes, yet already shows characteristic right-skewed shapes indicating fundamental career dynamics. Despite high variability in fit parameters, the distributions maintain consistent Weibull characteristics, with clear signatures of historical events visible in the year-to-year variations.

\subsubsection{Transition Period (1875-1924)}
As sample sizes grew, distribution shapes stabilised, marked by the emergence of consistent Weibull parameters ($k \approx 0.2$). This period witnessed reduced sensitivity to external events and the development of heavy tails in the distributions, suggesting the establishment of long-term academic careers as a stable phenomenon.

\subsubsection{Modern Period (1925-1960)}
The modern era shows highly stable distributions with clear Weibull characteristics, maintaining a consistent shape parameter $k \approx 0.2$ while demonstrating an increasing scale parameter $\lambda$. This combination indicate systematically longer average career durations, with robust power-law tails suggesting well-established career paths and institutional structures.

\subsection{Entertainment Industry Career Patterns}

The Internet Movie Database (IMDb) network presents distinctly different career patterns from academia. Entertainment careers show a higher Weibull shape parameter ($k \approx 0.5$), indicating different career termination dynamics. These distributions maintain remarkable stability across time periods and show less pronounced historical phase transitions. While entertainment careers typically have shorter average durations than academic ones, they exhibit more predictable termination patterns.

\subsection{Collaboration Dynamics Across Domains}

The evolution of collaboration patterns reveals fundamental differences between academic and entertainment networks:

\subsubsection{Academic Collaborations}
Academic collaborations in the MAG network show a clear three-phase evolution paralleling career patterns. Early-period collaborations exhibit exponential-like decay with high sensitivity to external events. The transition period (beginning around 1900) marks the emergence of heavy-tailed distributions and stable power-law exponents. The modern period shows remarkable stability, with consistent power-law tails (exponents between $2.1$ and $2.4$) and the Weibull parameters indicate institutionalised collaboration patterns. A clear separation exists between short-term and long-term collaboration regimes.

\subsubsection{Entertainment Industry Collaborations}
IMDb collaboration distributions maintain consistent characteristics throughout the network's history, with stable power-law exponents around $2.6$. Entertainment collaborations typically have shorter durations than academic ones and show notably less sensitivity to historical events, suggesting more resilient industry structures.

\subsection{Universal Features and Cross-Network Comparisons}

Despite their differences, both networks exhibit several universal features that suggest fundamental similarities in how professional careers evolve. A key characteristic shared across both domains is the maintenance of shape parameters $k < 1$, indicating that ``early failure'' career patterns represent a fundamental feature of professional collaboration networks. Additionally, both networks show persistent power-law tails across all time periods, with Weibull distributions providing consistently good fits despite dramatic network growth. Within each network, shape parameters maintain remarkable stability even as scale parameters evolve differently, reflecting the distinct characteristics of their professional environments.

The temporal evolution of fitting parameters reveals fundamental aspects of network development. A clear divergence emerges: while academic collaboration durations steadily increase, entertainment industry durations maintain stable characteristics. This divergence accelerates notably after 1950, even as goodness-of-fit metrics improve over time for both networks.

These findings provide insights into network evolution. The consistency of distribution shapes despite substantial network growth may indicate underlying social mechanisms. The persistent distinctions between academic and entertainment networks, particularly in their response to historical changes, show the strong influence of institutional structures. The universal applicability of Weibull distributions to career length modelling, despite these differences, points to common mechanisms underlying professional career dynamics across diverse fields.

\section{Discussion}
\label{sec:discussion}

Our analysis reveals fundamental patterns in how careers and collaborations evolve across different professional domains, with implications for our understanding of social network formation and professional development. We organise our discussion around three key findings: the universal characteristics of career longevity, the divergent evolution of collaboration patterns, and the distinct responses to external influences across academic and entertainment networks.

\subsection{Universal Patterns in Career Longevity}

The persistence of Weibull distributions in career duration across both networks suggests fundamental mechanisms underlying professional longevity. Our finding of shape parameters $k < 1$ in both domains (approximately $0.2$ for academic and $0.5$ for entertainment careers) shows that ``early failure'' represents a universal feature of collaborative careers. This aligns with recent findings by \citet{petersen2011} in scientific careers and \citet{yang2016} in artistic pursuits, suggesting these patterns may be more fundamental than previously recognised.

The stability of these parameters across centuries of evolution and dramatic institutional changes extends our understanding of career dynamics. While \citet{sinatra2016} demonstrated universal patterns in peak creative productivity, and \citet{liu2021} showed similar early-career vulnerability across scientific disciplines, our results show that these patterns persist across vastly different professional contexts and historical periods. This presents an alternative perspective to \citet{way2019}'s observation that early-career outcomes appear more stochastic in modern contexts—-- while individual outcomes may appear random, the underlying statistical patterns remain remarkably consistent.

\subsection{Divergent Evolution of Collaboration Patterns}

The distinct evolution of collaboration patterns between academic and entertainment networks illuminates how universal mechanisms interact with domain-specific factors. The steady increase in academic collaboration durations (shown by rising $\lambda$ values) contrasts with the stability of entertainment industry collaborations. This divergence supports recent observations by \citet{wu2019} regarding the increasing role of large teams in science, while revealing deeper historical patterns.

The remarkable stability of entertainment industry collaboration patterns, despite massive industry growth, extends findings by \citet{uzzi2013} on team formation dynamics in creative fields. Our longer temporal span shows this stability reaches much further back than previously documented. These contrasting patterns suggest that while career longevity may follow universal statistical laws, collaboration dynamics are strongly shaped by institutional and professional contexts.

\subsection{Institutional Mechanisms and Network Evolution}

The contrasting evolution of collaboration patterns between academic and entertainment networks reflects fundamental differences in their institutional structures. In academia, the steady increase in collaboration duration (power-law index evolving from 1.6 to 2.3) coincides with the post-World War II transformation of research funding structures \citep{price1976, allison1982}. The rise of multi-year grant programs created strong institutional incentives for sustained research partnerships, while departmental organization provided infrastructure supporting persistent collaborations through shared resources and facilities \citep{fountain2022}. This institutional framework, combined with tenure systems that reward sustained research output \citep{way2019}, has systematically encouraged longer-term collaborative relationships.

The entertainment industry's more stable collaboration patterns (index remaining between 2.6 and 2.1) emerge from distinctly different institutional mechanisms. While the industry has evolved from the classical studio system to more flexible production arrangements \citep{wasserman2018}, its fundamental project-based organization has remained consistent. Professional organizations and guilds have maintained stable working conditions throughout this evolution, while the intermediary role of talent agencies in structuring collaborations has persisted \citep{fraiberger2018}. These institutional structures appear to naturally generate more ephemeral but regularly renewed collaborative relationships, despite supporting similar career longevity patterns to academia.

The consistency of these collaborative patterns during network growth suggests institutional frameworks may play a more significant role than scale effects in shaping collaboration dynamics. This finding extends recent work by \citet{wang2023} on universal features of scientific collaboration networks, demonstrating how domain-specific institutional structures can maintain distinct collaborative patterns even while preserving universal career longevity distributions. Future research should examine how emerging institutional forms, particularly in digital and hybrid environments, might modify these established patterns. Understanding these institutional mechanisms has significant implications for policy design, suggesting that funding agencies and industry regulators should carefully consider how their interventions either reinforce or modify existing collaborative dynamics \citep{ghosh2022}.

\subsection{Limitations and Future Directions}

While our datasets span unprecedented temporal ranges, they represent only two types of professional collaboration networks. Future research should examine these patterns in other professional contexts, particularly those with different institutional structures. Higher temporal resolution data could reveal finer-scale dynamics that our yearly aggregated data cannot capture, while cross-cultural comparisons could illuminate how different social contexts influence these patterns.

The relationship between network size and environmental sensitivity deserves particular attention. Our findings suggest complex interactions between network scale and resilience, but the mechanisms underlying these relationships remain unclear. Understanding how institutional structures influence parameter stability could provide practical insights for organisational design.

\subsection{Practical Implications}

The universal aspects of career length distributions, combined with domain-specific evolution of collaboration patterns, have significant implications for institutional design and professional development. The consistent ``early failure'' patterns suggest the need for targeted early-career support systems, while the stability of collaboration patterns in entertainment might offer lessons for building resilient professional networks in other domains.

Understanding these principles could help institutions develop more effective strategies for fostering productive collaborations while supporting individual career development. The asymmetric response to external perturbations suggests that maintaining collaborative structures during disruptions might be as important as supporting individual careers.

\section{Conclusions}

Our comprehensive analysis of the Microsoft Academic Graph (1800-2020) and Internet Movie Database (1900-2020) reveals fundamental patterns in how professional collaboration networks evolve. These patterns suggest potential refinements to existing theoretical frameworks while offering new perspectives.

A key finding emerges in the observed consistency of career duration patterns. Both academic and entertainment networks show consistent Weibull distributions governing career longevity, with shape parameters that remain stable despite centuries of institutional and technological change. The academic network maintains $k \approx 0.2$, while the entertainment network shows $k \approx 0.5$. These values, consistently below 1, show that early career vulnerability represents a universal feature of professional collaboration networks. This finding transcends specific institutional contexts and historical periods, suggesting fundamental social mechanisms shape career trajectories in collaborative environments.

The evolution of collaboration patterns tells a nuanced story of universal features interacting with domain-specific characteristics. While the academic network shows systematically increasing collaboration durations (power-law index evolving from $1.6$ to $2.3$), the entertainment network maintains more stable patterns (index decreasing from $2.6$ to $2.1$). This divergence occurs even as both networks maintain stable career length distributions, showing that while career longevity mechanisms may be universal, collaboration dynamics remain strongly influenced by institutional contexts.

These findings suggest that current theoretical models of network evolution require significant revision. Future frameworks should incorporate Weibull-distributed removal processes as a fundamental feature rather than an emergent property. These models need to capture both the domain-specific evolution of collaborations and the asymmetric response to environmental perturbations, while explaining how networks maintain structural stability despite accelerating growth.

The persistence of these patterns through centuries of dramatic institutional and technological change suggests they reflect fundamental aspects of human collaborative behaviour. Understanding these principles holds significant implications for institutional design, career development programs, and the fostering of productive collaborations across professional domains.

Future research should extend these investigations to other types of professional networks while developing theoretical frameworks that can capture both the universal features and domain-specific variations we observe. The increasing availability of high-resolution digital datasets offers opportunities to examine these dynamics at finer temporal scales. Meanwhile, historical records from other professional domains could help establish the generality of these patterns. Integrating these quantitative findings with a qualitative understanding of institutional changes and career development could illuminate the mechanisms underlying the remarkable stability we observe in these evolving social systems.

\begin{acks}
This research was carried out at Rinna K.K., Tokyo, Japan.
\end{acks}
\bibliographystyle{SageH}
\bibliography{paper2}

\end{document}